\documentclass[12pt]{article}
\usepackage{amsmath}
\usepackage{amsfonts}
\usepackage{graphicx,psfrag,epsf}
\usepackage{enumerate}
\usepackage[round]{natbib}
\usepackage[usenames, dvipsnames]{color}
\usepackage{url} 
\usepackage[noblocks]{authblk}

\usepackage{mdframed} 
\usepackage{caption}
\usepackage[labelformat=simple]{subcaption}

\usepackage{siunitx}
\usepackage{float}
\usepackage[normalem]{ulem}
\newcommand{\stkout}[1]{\ifmmode\text{\sout{\ensuremath{#1}}}\else\sout{#1}\fi}
\usepackage{mathtools}
\usepackage{mdwlist}

{
\begin{basedescript}{
\desclabelwidth{2em}}}
{
\end{basedescript}
}

\newcommand{\blind}{0}
\usepackage[table,xcdraw,dvipsnames]{xcolor}
\usepackage{diagbox}
\usepackage{bbm}
\usepackage{algorithm}
\usepackage[noend]{algpseudocode}

\usepackage{multirow}
\usepackage{adjustbox}  

\usepackage{inputenc}
\usepackage[english]{babel}
\usepackage{amsthm}
\usepackage{mathtools}

\usepackage{hyperref}
\hypersetup{
    colorlinks=true,
    linkcolor=blue,
    citecolor=blue,
    filecolor=magenta,      
    urlcolor=blue,
}

\newcommand{\bs}{\boldsymbol{s}}

\newcommand{\numStations}{438}
\newcommand{\btheta}{\boldsymbol{\theta}}

\newcommand{\degree}{$^\circ$}

\makeatletter
\renewcommand*\env@matrix[1][\arraystretch]{%
  \edef\arraystretch{#1}%
  \hskip -\arraycolsep
  \let\@ifnextchar\new@ifnextchar
  \array{*\c@MaxMatrixCols c}}
\makeatother

\begin{document}
\if0\blind
{\title{\bf Explaining the unexplainable: leveraging extremal dependence to characterize the 2021 Pacific Northwest heatwave}
\date{}
\author[1,2]{Likun Zhang}
\author[1]{Mark D. Risser}
\author[1]{Michael F. Wehner}
\author[3,1]{Travis A. O’Brien}
\affil[1]{Climate and Ecological Sciences Division, Lawrence Berkeley National Laboratory, USA
}
\affil[2]{Department of Statistics, University of Missouri, Columbia, MO}
\affil[3]{Department of Earth and Atmospheric Sciences, Indiana University, Bloomington, IN, USA}
\maketitle
}\fi

\if1\blind
{
{\title{\bf Explaining the unexplainable: leveraging extremal dependence to characterize the 2021 Pacific Northwest heatwave}}
\author{}
\date{}
\maketitle
}\fi


\begin{abstract}
   In late June, 2021, a devastating heatwave affected the US Pacific Northwest and western Canada, breaking numerous all-time temperature records by large margins and directly causing hundreds of fatalities. The observed 2021 daily maximum temperature across much of the U.S. Pacific Northwest exceeded upper bound estimates obtained from single-station temperature records even after accounting for anthropogenic climate change, meaning that the event could not have been predicted under standard univariate extreme value analysis assumptions. In this work, we utilize a flexible spatial extremes model that considers all stations across the Pacific Northwest domain and accounts for the fact that many stations simultaneously experience extreme temperatures. Our analysis incorporates the effects of anthropogenic forcing and natural climate variability in order to better characterize time-varying changes in the distribution of daily temperature extremes. We show that greenhouse gas forcing, drought conditions and large-scale atmospheric modes of variability all have significant impact on summertime maximum temperatures in this region. Our model represents a significant improvement over corresponding single-station analysis, and our posterior medians of the upper bounds are able to anticipate more than 96\% of the observed 2021 high station temperatures after properly accounting for extremal dependence. 
\end{abstract}

\noindent%

\section{Introduction} \label{sec:intro}
The immediate synoptic causes of the deadly and unprecedented heatwave across the US Pacific Northwest (PNW) in 2021 are quite clear. An eastward moving wave train propagated from the western subtropical Pacific and developed into an \emph{omega} 
atmospheric blocking pattern centered over Western Canada \citep{McKinnon_Simpson}. Here, the term \emph{omega} alludes to the shape of the mid-latitude jet stream, a narrow, persistent band of high winds aloft, during such meteorological conditions, with relatively warm air and high pressure occurring at the surface under the northward bend of the jet stream. Popularly known as a `heat dome', the resulting local high pressure and associated anticyclonic circulation caused downward air motion and heating by adiabatic compression \citep{Wang2022}, while clear skies enhanced solar heating at the surface \citep{Mo2022}.  Such meteorological conditions are common during heatwaves, however this one differed from known cases in that the air above the Pacific Northwest was pre-heated by condensation (clouds and precipitation) within an atmospheric river that occurred several days prior over the Pacific Ocean. This pre-heated air then rose further in temperature as it descended toward the surface and compressed due to increasing pressure (adiabatic compression). In short, an unusual summer trans-Pacific atmospheric river injected latent heat energy and moisture raising temperatures further \citep{Mo2022}. Furthermore, the intensity of the blocking pattern may been influenced by an Arctic polar vortex split preceding the heatwave \citep{Wang2022}. Despite this complicated meteorology, the unprecedented temperatures and other heat indices experienced during the 2021 PNW heatwave were well predicted several days in advance by forecast centers \citep{Emerton_et_al}. Undoubtedly, anthropogenic climate change also played a role in the 2021 PNW heatwave but its far outlier temperatures challenge extreme value statistical analyses to confidently predict the event's rarity from previous observations \citep{emily2022dynamics,Philip21}. In these two previous studies, the upper bound of a Generalized Extreme Value (GEV) distribution fitted to observed maximum summer daily maximum temperatures, from station data prior to 2021, was lower than the hottest observed 2021 temperatures. As a result, it is impossible to explain the true abnormality of this event or quantify the probability of a similar event in the future using their nonstationary extreme value statistical methods. 

Alternatively, \citet{emily2022dynamics} performed a storyline hindcast attribution study using two regional climate models (the Weather Research and Forecasting model, WRF, and RegCM) to attempt to quantify the magnitude of the human influence without addressing the question of the changes in rarity. 
Their study found that human-induced climate change increased the 2021 PNW heatwave temperature by approximately 1\degree C. 
\citet{Philip21} included the 2021 temperatures in a fitted GEV distribution and used that information along with summer annual maximum temperatures from long climate model simulations to find a human increase in the 2021 PNW heatwave temperature of about 2\degree C. Based on this in-sample GEV analysis, they further concluded that the event was ``virtually impossible'' without anthropogenic climate change. However, \citet{emily2022dynamics} found that including the 2021 temperatures decreases the $p$-values of the $\chi^2$ goodness-of-fit tests from greater than 0.2 in general to less than 0.05 at many stations, implying that the 2021 seasonal maxima are generated from a different underlying distribution than the previous years. \citet{McKinnon_Simpson} used a large climate model ensemble simulation to examine gridboxes across the global land where simulated maximum temperatures exhibited similar skewness and kurtosis to the observed station temperatures in the PNW region. They found that while the model could produce temperatures that exceed 4.5 standard deviations above average temperatures (similar to the 2021 PNW observations), such occurrences are extremely rare with return periods of about 100,000 years. Finally, \citet{Heeter2023} examined tree ring records finding that the 2021 summer temperatures in the PNW regions were unprecedented in the last millennium.

In this paper, our objective is to examine the statistical difficulties of estimating the probability of far outlier temperatures along with human- and naturally-driven changes to such quantities. Such estimates are required to make the attribution statements that advance our understanding of how anthropogenic forcing and natural climate variability affected the rarity of the PNW heatwave, similarly to those in \citet{Philip21} and \citet{emily2022dynamics}. The primary challenge is 
that the out-of-sample GEV analyses in these two studies both yielded estimated upper bounds for temperature extremes lower than the observed 2021 daily maxima, meaning the PNW heatwave was impossible to anticipate in advance of its occurrence under station data-based models. Consequently, the human influence on these ``impossible'' events cannot be attributed, either in likelihood or severity because they are not within the bounds (i.e., supports) of the out-of-sample model fits. This unexplainability is largely due to the fact that analyses in the two studies represent simplified approaches to the problem of estimating the spatial extent of an event like the PNW heatwave: \citet{Philip21} analyzed the spatial average of annual maxima from all weather stations in a fixed latitude-longitude box (which both aggregates over important topographical variability in the PNW domain and ignores the fact that historical events may have been offset geographically from this box), while \citet{emily2022dynamics} utilize single-station analyses (which ignore both autocorrelation in the climatology of temperature extremes and the fact that individual heatwave events will impact multiple weather stations). As outlined in \citet{zhang2022storm}, failing to account for the spatial coherence of individual events 
results in underestimation of the risks associated with weather extremes \citep{davison2012statistical, saunders2017spatial}. 
Our hypothesis is that extreme value analysis methods that more appropriately account for both climatological- and event-based-autocorrelation \citep[e.g.,][]{huser2017bridging,huser2019modeling,zhang2021hierarchical} can better characterize temperature extremes in the PNW such that the 2021 temperatures are less ``unexplainable'' in an out-of-sample statistical sense. Importantly, these methods should also improve the predictive power of the out-of-sample analysis on the 2021 PNW heatwave so that we can construct well-defined attribution statements.

To test this hypothesis, we analyze the summertime maxima of homogenized Pacific Northwest U.S. temperature station data from 1950 to present using a statistical model that flexibly accounts for spatial variability in the climatology of extreme temperatures and yields data-driven measures of the spatial extent of heatwave events in the PNW. Our statistical framework can account for both secular trends due to anthropogenic and naturally forced year-to-year variability while robustly quantifying uncertainty via Bayesian hierarchical modeling. We explicitly quantify the individual effect of including physical covariates versus accounting for the spatial coherence of heatwave events on the corresponding GEV upper bounds and whether or not the 2021 temperature extremes could have been anticipated in an out-of-sample sense. We then revisit the climate change attribution question, using a statistical counterfactual \citep[following][]{risser2017attributable} invoking Granger causal inference \citep{granger1969investigating} to isolate the effect of anthropogenic forcing on both the predictability of the 2021 temperatures and their probabilities.

The paper proceeds as follows. Section \ref{sec:data} describes all input data sources for our analysis, 
while in Section \ref{sec:methods} we outline our methodological innovations, including the construction of the Bayesian hierarchical model (Section \ref{sec:eva}), an attribution framework for isolating the human influence on the heatwave 
(Section \ref{subsec:statCF}), and descriptions of alternative statistical models 
(Section \ref{subsec:altMods}). Our main results are provided in Section \ref{sec:results}, and Section \ref{sec:conc} concludes the paper.




\section{Data}\label{sec:data}
\subsection{Long-term temperature records for trend analysis}
In order to assess secular trends over time and year-to-year variability that are unaffected by inhomogeneities from nonclimatic influences (such as changes to measurement devices and the local effects of the built environment), we analyze a homogenized version of the Global Historical Climatology Network-Daily (GHCN-D) over the contiguous United States \citep{rennie2019development}. While the homogenized records extend back to the early 20th century, we focus our analysis on the last 71 years (1950-2020) due to the high density of nonmissing measurements from weather stations during this period. Focusing our attention on the 487 gauged locations from this database within a longitude-latitude box defined by $[125^\circ\text{W},116.5^\circ\text{W}]\times[40^\circ\text{N},49^\circ\text{N}]$ (see Supplemental Figure~\ref{fig:topoCovt}(b)), we extract the summertime (June/July/August or JJA) maximum daily maximum temperature measurement, denoted ``TXx,'' from each station in each year so long as there are no more than 40\% missing daily measurements in each JJA season. Stations with less than ten non-missing JJA TXx from 1950-2020 are excluded; this results in $n=438$ gauged locations with sufficient temperature records. The seasonal maxima from these locations are used in the statistical analysis described in Section~\ref{sec:methods}; note that we do not include any measurements from 2021 in our analysis of the historical record (i.e., the 2021 event is considered out-of-sample).

\begin{figure}
\centering
\captionsetup[subfigure]{justification=centering}
    \begin{subfigure}[t]{.425\textwidth}
\centering
\caption{2021 TXx}
\includegraphics[width=\textwidth]{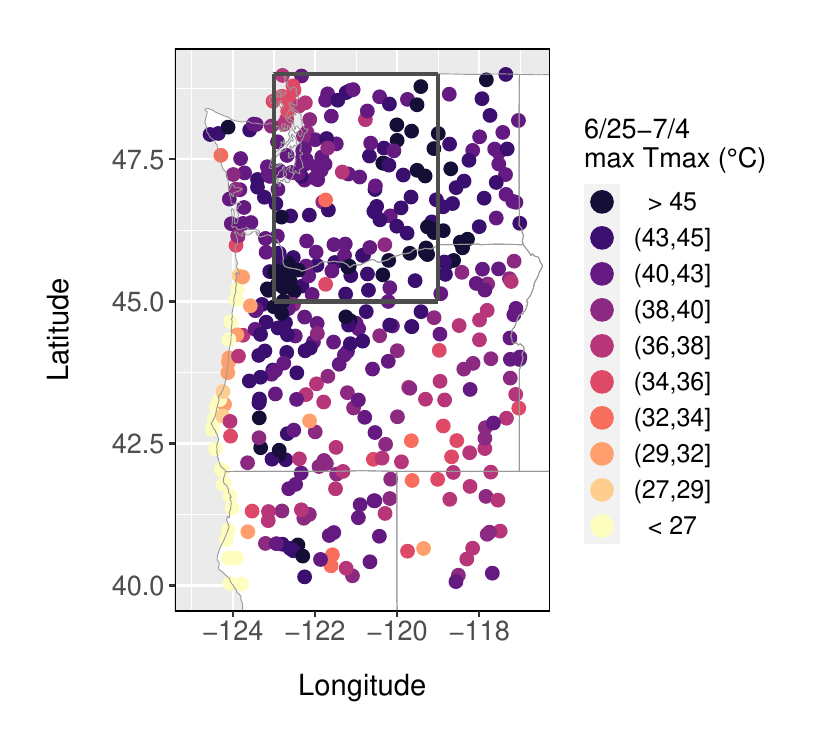}
    \label{subfig:maxTmax2021}
    \end{subfigure}
    \begin{subfigure}[t]{.38\textwidth}
\centering
\hspace*{1em}\caption{Area-averaged JJA TXx}
\includegraphics[height=\textwidth]{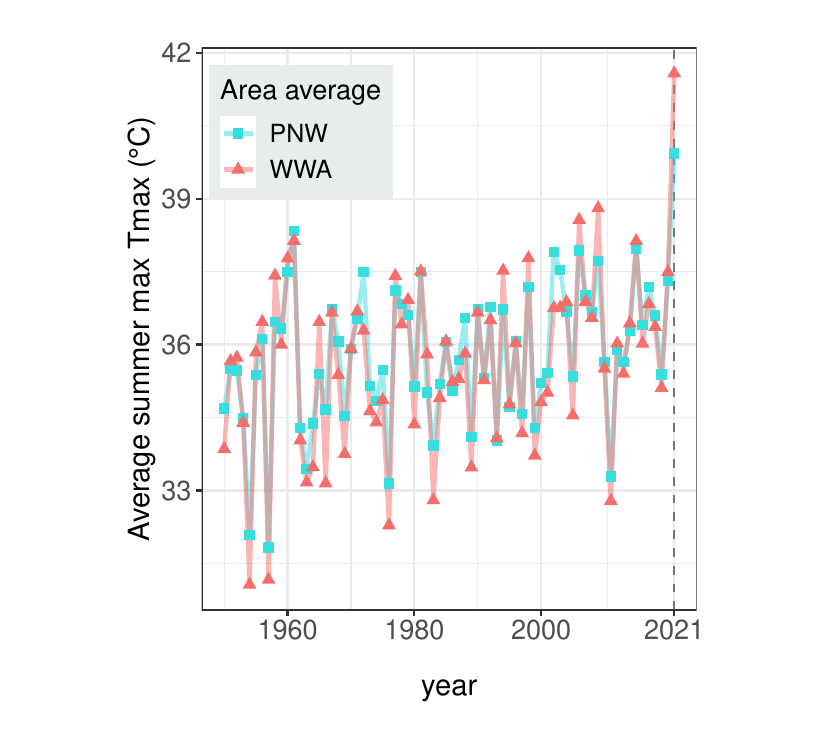}
    \label{subfig:avg_timeseries}
    \end{subfigure}
\vspace*{-2.7em}
\caption{
    \subref{subfig:maxTmax2021} The maximum daily maximum temperature between June 25 and July 4 (i.e., TXx) at each of the $n=470$ high-quality Global Historical Climatology Network gauged locations in the Pacific Northwest region with the World Weather Attribution \citep[WWA;][]{Philip21} region 
    overlaid, 
    \subref{subfig:avg_timeseries} area-averaged June/July/August (JJA) TXx time series from 1950-2021 for the PNW and WWA regions.
    } 
    \label{fig:data_covariates}
\end{figure}

\subsection{Estimation of the 2021 PNW temperature extremes}

Homogenized temperature records are used for trend analysis because they explicitly modify the raw station measurements to account for known, non-natural shifts in the distribution of daily temperatures. However, to quantify the extreme daily temperatures that actually occurred during the heatwave period (defined as June 25 to July 4, 2021), we prefer to use the available non-homogenized GHCN-D \citep{Menne2012} records from the longitude-latitude box shown in Figure~\ref{fig:data_covariates}. From the GHCN-D records with available daily maximum temperatures over these ten days, we see that a large majority of the maximum daily measurements (the 2021 TXx) occurred between June 26 and July 1 (see Supplemental Figure~\ref{fig:dateOfTXx}); there are $n=470$ GHCN-D gauged locations with six non-missing daily maximum temperatures over these six days (see Supplemental Figure~\ref{fig:GHCNcode}).
Unfortunately, there is a mismatch in the $n = 438$ homogenized sites used to estimate historical trends and the $n = 470$ GHCN-D sites used to define the 2021 event
(see Supplemental Figure~\ref{fig:compareTXx}a.).

To ensure that we can report results for all gauged locations in the homogenized records, we apply the spatial statistical model proposed in Eq.~\eqref{eq:TPS}, wherein we use thin plate splines and topographical covariates to conduct interpolation to a regular 800m grid over the PNW domain. Supplemental Figure~\ref{fig:pnw_TXx} shows the GHCN-D TXx used as input data as well as the results of this interpolation for the $n=470$ GHCN-D sites (Supp. Fig~\ref{fig:pnw_TXx}(b)), the $n=438$ homogenized sites (Supp. Fig~\ref{fig:pnw_TXx}(c); these values are identical to the ones shown in Figure~\ref{fig:data_covariates}), and the 800m grid (Supp. Fig~\ref{fig:pnw_TXx}(d)). As shown in Supplemental Figure~\ref{fig:compareTXx}(b) and (c), the interpolated TXx measurements compare favorably with the GHCN-D input data ($R^2=0.94$), and furthermore provide a better reconstruction than corresponding measurements from the nearest-neighbor ERA5 grid box \citep[ERA5 is a state-of-the-art atmospheric reanalysis product,][that is widely used in climate analysis]{hersbach2020era5,bell2021era5}.

\subsection{Anthropogenic and natural drivers of temperature change}\label{sec:drivers}

One benefit of our methodology is that we can include physical covariates to account for spatio-temporal nonstationarity in GEV distributions to describe both long-term trends in the distribution of temperature extremes as well as year-to-year variability due to modes of natural climate variability (see Section~\ref{sec:methods}). 

\paragraph{Secular trends.} Dating back to seminal work by \cite{arrhenius1897influence}, it is well-established that anthropogenic greenhouse gas (GHG) emissions drive increases to global mean temperature via radiative forcing of the climate system; this was most recently underscored in the Intergovernmental Panel on Climate Change (IPCC) Sixth Assessment Report, which stated ``Human activities, principally through emissions of greenhouse gases, have unequivocally caused 
global warming'' \citep{Lee2023}. Following \cite{Risser2022}, we use the sum-total forcing time series of the five well-mixed greenhouse gases (CO$_2$, CH$_4$, N$_2$O, and the CFC-11 and CFC-12 halocarbons) to describe long-term, human-induced secular trends in the distribution of extreme temperature measurements. The radiative forcing time series (see Figure \ref{subfig:GHG}) is calculated from reconstructed atmospheric concentrations of each greenhouse gas \citep{meinshausen2016, meinshausen2018} via the forcing formulae given in \cite{Etminan2016radiative} and \cite{hodnebrog2013global}; a time lag is also included to account for the lagged response of sea surface temperatures to these forcings \citep[see][for further details]{Risser2022}. Thus, the greenhouse gas forcing time series, denoted by $\text{GHG}_t$, $t=1950,\ldots, 2021$, characterizes a nonlinear trend over 1950-2021, and is applied uniformly across space.

\paragraph{Large-scale modes of oceanic variability.} The El Ni\~no-Southern Oscillation (ENSO) is a 
coupled ocean-atmosphere   mode of natural climate variability that cycles between positive (El Ni\~no) and negative (La Ni\~na) phases every two to seven years \citep{Philander1985}. While the effect of ENSO on winter climate in the Pacific Northwest is robust \citep{Whan2017}, ENSO also influences the large scale circulation in the region during the summer \citep{Stuecker2015}. Given the complicated dynamics responsible for the 2021 PNW event, we use the ENSO Longitude Index \citep[ELI,][]{Williams2018diversity} as a covariate to represent this influence.  ELI is a sea surface temperature-based index that summarizes the average longitude of deep convection in the Walker Circulation 
and accounts for the different spatial patterns of ENSO. The JJA seasonally-averaged ENSO time series is applied uniformly across space and is denoted by $\text{ELI}_t$, $t=1950,\ldots, 2021$; see Figure \ref{subfig:ELI}.

\paragraph{Palmer Drought Severity Index.} Heatwave temperatures are strongly influenced by  evapotranspirative cooling from the surface soil moisture content and the local vegetation \citep{Domeisen2023}. Furthermore, high temperatures can reduce this available surface moisture. To capture this feedback in our statistical model we use the Palmer Drought Severity Index (PDSI) as another physically motivated covariate \citep{palmer1965meteorological}. PDSI quantifies the relative surface dryness through moisture supply and demand by calculating the cumulative departure in water balance using precipitation and temperature data. In our analysis, we use the PRISM monthly PDSI data from the West Wide Drought Tracker \citep{abatzoglou2017west}, and flip the sign of their indices so that PDSI ranges from about -10 (wet) to +10 (dry) with values above 3 representing severe to extreme drought. We then calculate the JJA average for each year $t=1950,\ldots, 2021$ at each spatial location $\bs$ (denoted by $\text{PDSI}_t(\bs)$). Just prior to the 2021 PNW heatwave, much of Washington State and Oregon were in preexisting severe drought conditions (see Figure \ref{subfig:pdsi}).

\paragraph{Urbanization Binary Index.} The urban heat island effect is another well-known anthropogenic factor influencing temperature distributions within urban and suburban environments \citep[e.g.,][]{Chen2019} due to differences in heat fluxes (both sensible and latent) and momentum fluxes (e.g., surface roughness) that impact the evolution of daily temperatures \citep{Oleson2011}. To account for the effects of urbanization in a parsimonious manner, we treat urbanization as a binary covariate.  \citet{markley2022housing} developed a reliable ``urbanization year''  indicator, based on United State census data, that denotes if and when census tracts became ``urbanized'' during 1940-2021. They define a region as urbanized if its housing density exceeds 200 housing units per square mile. To model the potential urban heat island effect, we use the dataset of \citet{markley2022housing} to create a binary variable $\text{Urban}_t(\bs)$ for each year $t=1950,\ldots, 2021$ to indicate whether the census tract surrounding the station $\bs$ has been urbanized in year $t$ by finding the corresponding census tract and its urbanization year.

\subsection{Characterizing spatial nonstationarity in GEV climatology} \label{sec:spatNonstat}

The anthropogenic and natural drivers described in Section~\ref{sec:drivers} are used to account for nonstationarity over space and time 
in the distribution of extreme temperature measurements (see Eq.~\eqref{eqn:gev_coefs}). However, we use an additional set of covariates to statistically model spatial nonstationarity in the parameters of the extreme temperature distributions (described in Eq.~\eqref{eq:TPS}). In addition to thin plate splines, the following spatial-only covariates are used to statistically model spatial variability in distributional parameters:
\begin{itemize}
    \item The natural logarithm of the average monthly precipitation over January, 1950, to December, 2020, as estimated from the PRISM data set \citep{daly1994statistical}. This covariate is used to discriminate between wet and dry regimes of the PNW.
    \item Topographical aspect, slope, and elevation. Elevation (meters above sea level) accounts for the fact that extreme temperatures tend to decrease with increasing altitude (via lapse rate theory). Topographical aspect describes the direction in which each location faces (ranging from zero to $2\pi$), while topographical slope is an index that describes the steepness; collectively, slope and aspect account for the effect of solar radiation and exposure on extreme temperatures. Elevation is obtained from the 800m digital elevation map used to generate the PRISM data product; slope and aspect are calculated using the \texttt{raster} package for \textbf{R} \citep{R_raster}.
    \item Distance-to-coast (km) to account for the fact that stations closer to the ocean tend to have cooler temperatures. This variable is calculated by \citet{dist2coast} at a global grid of $1/16^\circ$ using the Generic Mapping Tools package \citep{wessel1998new}. They did not consider the landlocked bodies of water (e.g., Lake Tahoe) to be oceans with a coast.
\end{itemize}
Maps of each of these quantities are shown in Supplemental Figure~\ref{fig:topoCovt}. Since a $1/16^\circ$ degree grid is very high resolution, we simply interpolate them to the GHCN locations to obtain the values for model fitting.

\section{Methods} \label{sec:methods}

\subsection{Spatial extreme value analysis} \label{sec:eva}

In order to assess the effect of statistical assumptions on the failure of previous 
studies to characterize a nonzero probability in an out-of-sample context (as described in Section~\ref{sec:intro}), we first describe a more refined nonstationary statistical model that flexibly accounts for spatial coherence in both the climatology of temperature extremes and individual heatwave events. The methodology is grounded in extreme value theory and is comprised of three building blocks, each of which represents important innovations relative to existing observational analyses of the 2021 PNW heatwave, and each of which is specifically tailored to the statistical modeling of daily temperature extremes in PNW: (1) nonstationary marginal GEV analysis including multiple covariates, (2) accounting for climatological spatial dependence, and (3) accounting for spatial coherence among individual events. While presented separately in the following paragraphs, these components are incorporated within a single Bayesian  hierarchical model. 
All unknown statistical parameters are assigned noninformative prior distributions, and Markov chain Monte Carlo methods are used to generate samples from the resulting posterior distribution, upon which all subsequent inference is based. We now describe each of these components in turn.

\paragraph{Nonstationary marginal Generalized Extreme Value analysis.} The first innovation of our methodology involves incorporating of a broad range of external information (also known as ``covariates'') to characterize spatially- and temporally-varying changes in the estimated distribution of extreme temperature.
Based on classical extreme value theory \citep[see Theorem 3.1.1 of][]{Coles2001}, the JJA maxima in year $t$ at a generic weather station $\bs$ from PNW, denoted by $Y_t(\bs)$, can be considered as arising from a generalized extreme value (GEV) distribution with the cumulative distribution function (CDF) 
\begin{equation}\label{eqn:gev_fam}
\mathbb{P}(Y_t(\bs) \leq y) = \exp\left\{-\left[ 1 + \xi_t(\bs)\left(\frac{y - \mu_t(\bs)}{\sigma_t(\bs)}\right) \right]^{-1/\xi_t(\bs)} \right\}, 
\end{equation}
which is defined for $\{ y: 1 + \xi_t(\bs)(y - \mu_t(\bs))/\sigma_t(\bs) > 0 \}$. The GEV distribution is characterized by the location parameter $\mu_t(\bs) \in \mathbb{R}$, 
the scale parameter $\sigma_t(\bs)>0$ 
and the shape parameter $\xi_t(\bs) \in \mathbb{R}$. 
While the three GEV parameters provide useful information about the marginal behavior of temperature extremes over time, 
we are often more interested in summaries of the fitted GEV distribution. For the purposes of this analysis, we are interested in the so-called risk probability for the 2021 heatwave, 
which summarizes the probability of exceeding the location-specific TXx temperature threshold that occurred during the heatwave (see Figure~\ref{subfig:maxTmax2021}; denoted by $u(\bs)$) in a given year $t$. Based on the form of the GEV distribution, we can calculate the risk probability as a direct function of the GEV parameters \citep{Coles2001}:
\begin{equation} \label{eqn:riskProb}
{p}_t(\bs, u(\bs)) = \left\{ \begin{array}{ll}
1 - \exp\left\{- \left[1- {\xi}_t(\bs)\left[({\mu}_t(\bs) - u(\bs)\right]/{\sigma}_t(\bs) \right]^{-1/{\xi}_t(\bs)} \right\},  & {\xi}_t(\bs) \neq 0, \\[1ex]
1 - \exp\left\{- \exp\big\{ \left[({\mu}_t(\bs) - u(\bs)\right]/{\sigma}_t(\bs) \big\} \right\},  & {\xi}_t(\bs) = 0.
\end{array} \right. 
\end{equation}
Additionally, when the shape parameter $\xi_t(\bs)$ is negative, the GEV distribution has a finite, time-varying upper bound, which can also be written in terms of the GEV parameters as
\begin{equation} \label{eqn:upperBound}
b_t(\bs) = {\mu}_t(\bs) - {\sigma}_t(\bs)/\xi_t(\bs).
\end{equation}
The fact that the GEV distribution has a finite bound when $\xi_t(\bs)<0$ has important implications for analyzing temperature extremes; see Section~\ref{sec:results} for further discussion.

Specific to analyzing temperature extremes in the Pacific Northwest, the spatio-temporal variability in 
GEV parameters are modeled as:
\begin{equation}
    \label{eqn:gev_coefs}
\begin{array}{rcl}
    \mu_t(\bs) & = & \mu_0(\bs)+\mu_1(\bs)\text{GHG}_t + \mu_2(\bs)\text{PDSI}_t(\bs)  + \mu_3(\bs)\text{ELI}_t +\mu_4(\bs) \text{Urban}_t(\bs), \\
    \log \sigma_t(\bs) & = &  \sigma_0(\bs)+\sigma_1(\bs)\text{GHG}_t + \sigma_2(\bs)\text{ELI}_t, \\
    \xi_t(\bs) & \equiv & \xi(\bs),
\end{array}   
\end{equation}
in which $\text{GHG}_t$, $\text{PDSI}_t(\bs)$, $\text{ELI}_t$ and $\text{Urban}_t(\bs)$ are anthropogenic and natural drivers defined in Section \ref{sec:drivers}. We tested different combinations in a series of pointwise analyses and determined that this combination of covariates gives the maximum likelihood assuming independence among stations. Thus, $\{\mu_0(\bs), \dots, \mu_4(\bs), \sigma_0(\bs), \ldots, \sigma_2(\bs),  \xi(\bs)\}$ are called GEV coefficients, which describe the climatological behavior of temperature extremes. Statistically modeling spatio-temporal variability in the GEV parameters in this manner implies a nonstationary model for extremes, and furthermore means that the risk probabilities in Eq.~\eqref{eqn:riskProb} can be calculated as a function of the GEV coefficients and the drivers.

\paragraph{Accounting for climatological dependence.} The second important innovation of our statistical modeling framework accounts for the fact that the climatology of temperature extremes exhibits spatial dependence: two locations that are nearby are more likely to have similar climatological extremes than two locations that are far apart. This spatial coherence is accounted for 
by forcing the climatological GEV coefficients to vary smoothly over the geospatial domain of interest. To this end, we suppose each marginal GEV coefficient is a function of thin plate splines \citep{wood2003thin} and topographical features. In other words, for $\gamma\in\{\mu_0,\mu_1,\mu_2,\mu_3, \mu_4, \sigma_0,\sigma_1,\sigma_2,\xi\}$, we statistically model

\begin{equation} \label{eq:TPS}
\gamma(\bs)\vert\beta^{\gamma}_0,\beta^{\gamma}_1,\ldots, \beta^{\gamma}_B =  \beta^{\gamma}_0+\sum_{i=1}^{99} \beta^{\gamma}_i \underbrace{b_i(\bs)}_{\hbox{\tiny T.P. splines}} + \sum_{j=1}^{5} \alpha^{\gamma}_j \underbrace{g_j(\bs)}_{\hbox{\tiny topography}},
\end{equation}
where the thin plate splines $b_i(\cdot)$ are smooth functions of longitude and latitude and the $g_j(\bs), j = 1, \dots, 5$ are the spatial-only covariates described in Section~\ref{sec:spatNonstat} (log average precipitation, elevation, slope, aspect, and distance-to-coast; see Supplemental Figure~\ref{fig:topoCovt}). We included 99 thin plate spline functions in order to flexibly model fine-scale spatial variability; however, to guard against over-fitting we furthermore included a regularization prior for the $\{\beta^\gamma_i:i=1,\dots,99\}$ \citep[see, e.g.,][]{Hans2009}.

\paragraph{Accounting for the spatial coherence of individual events.} While enforcing spatial coherence in the GEV coefficients following Eq.~\eqref{eq:TPS} is useful for describing climatological dependence, 
we must further account for the fact that individual heatwave events (aggregated over the course of a JJA season) will simulatenously affect multiple gauged locations. The third innovation of our methodology is that we account for this so-called ``data-level'' dependence in the maxima using a flexible method from the spatial extremes literature called a Gaussian scale mixture; see \cite{zhang2021hierarchical} for a full treatment of the theory and \cite{zhang2022storm} for an application of the methodology to an analysis of extreme precipitation.
At each time $t$,  we assume the spatial field of JJA maxima $\{Y_t(\bs)\}$ has a copula defined by 
\begin{equation}\label{eqn:noisyScalemix}
    X_t(\bs)=R_t \cdot W_t(\bs)+\epsilon(\bs),
\end{equation}
where $R_t\sim\text{Pareto}\{(1-\delta)/\delta\}$, 
$W_t(\bs)$ is a standard isotropic and stationary Gaussian process with standard Pareto margins and $\epsilon(\bs)$ is a white Gaussian noise process with variance $\tau^2$. The covariance function of $W_t(\bs)$, $C_{\theta_W}(h)$, describes the covariance of $W_t(\bs)$ at pairs of locations separated by a distance $h$ and is indexed by $\theta_W=(\rho,\nu)$, in which $\rho$ is the range parameter
, and $\nu$ is the smoothness parameter that controls how rough or blotchy the copula looks. 
In addition, $R_t$ at time $t$ acts as a scaling factor that amplifies the simultaneous large but not yet extreme values in $\{W_t(\bs)\}$. In the meantime, larger $\delta$ induces heavier-tailed $R_t$ and thus the amplifying effect from $R_t$ will tend to be more evident. Consequently, the Gaussian scale mixture $\{X_t(\bs)\}$ exhibits asymptotic independence when $\delta\in (0,1/2]$ and asymptotic dependence when $\delta\in (1/2,1)$. Larger $\delta$ values also induce higher sub-asymptotic dependence strength, and in the case of $\delta\in (0,1/2]$ there is still weakening yet nonzero dependence at the sub-asymptotic levels. Using a well-defined dependence measure, we verified empirically in Appendix \ref{sec:chi_emp} that this copula model accurately quantifies the spatial dependence between stations during individual extreme heatwave events.

As described in Section \ref{sec:data}, we assume that marginally (i.e., at each station) the JJA maxima arise from the GEV distribution described by Eq.~\eqref{eqn:gev_fam} 
and~\eqref{eqn:gev_coefs}. Given that we have one measurement per year at each station, we assume that any long-term autocorrelation or year-to-year variability in the maxima at each station is fully described by the time-varying location and scale parameters and any residual variability is temporally independent. For simplicity, we henceforth write $X_{tj} = X_t(\bs_j)$, $\boldsymbol{X}_t=(X_{t1},\ldots,X_{t,\numStations})^T$ and so forth, in which $\bs_j$ denotes the location of $j$th station, $j=1,\ldots, \numStations$. In order to connect the copulas of the observed process $\{Y_t(\bs)\}$ and the Gaussian scale mixture $\{X_t(\bs)\}$, we define the following marginal transformation:
\begin{equation}\label{eqn:marginalTrans}
   F_{Y\vert\gamma(\bs),t}\{Y_t(\bs)\}=F_{X\vert\delta,\tau^2}\{X_t(\bs)\},
\end{equation}
where $F_{Y\vert\gamma(\bs),t}$ is the GEV distribution function with parameters $\{\mu_t(\bs), \sigma_t(\bs),\xi_t(\bs)\}$ defined in Eq.~\eqref{eqn:gev_coefs} and $F_{X\vert\delta,\tau^2}$ is the marginal CDF for $X_t(\bs)$ in Eq.~\eqref{eqn:noisyScalemix}. This marginal transformation 
is tied together with the copula \eqref{eqn:noisyScalemix} in a single Bayesian hierarchical model to make inference on the model parameters; see Supplemental Section \ref{sec:hierarchy_model} for more details.

\subsection{Statistical counterfactual for quantifying human influence} \label{subsec:statCF}

The statistical framework for extreme value analysis outlined in Eq.~\eqref{eqn:gev_fam} and \eqref{eqn:gev_coefs} uses covariates to describe long-term trends and year-to-year variability in temperature extremes in the Pacific Northwest, and as such we can use the fitted statistical model to isolate the human influence on various aspects of the heatwave following the approach used in \cite{risser2017attributable}. Specifically, the statistical model specified by Eq.~\eqref{eqn:gev_coefs} implies that risk probabilities and upper bounds (in the case of $\xi(\bs)<0$) are all functions of the input covariates, i.e., GHG forcing, the ELI, the PDSI, and the urbanization index. As such, we can use the fitted statistical model to estimate risk probabilities and upper bounds for arbitrary combinations of GHG forcing, ELI, PDSI, and the urbanization index, including combinations that did not actually occur over 1950-2021. Our results in Section~\ref{sec:results} involve comparing risk probabilities and upper bounds for two different climate scenarios (summarized in Table~\ref{tab:statCFs}): the ``factual'' quantities represent our best estimates for the conditions present when the heatwave occurred, and the ``counterfactual'' summarizes the \textit{isolated effect of GHG forcing} while holding all other quantities fixed.

\begin{table}[!t]
\caption{Summary of the two climate scenarios considered for quantifying human influence on aspects of the heatwave event and conducting attribution statements. The counterfactual is ``statistical'' in the sense that it is constructed from the fitted models using covariates (and correspond to conditions that never occurred in the real world). The years listed correspond to those used to determine the input values for the anthropogenic and natural drivers.}
\begin{center}
{\small 
\begin{tabular}{lcccc}
\hline\noalign{\smallskip}
 & \textbf{GHG forcing} & \textbf{Urbanization} & \textbf{ENSO longitude}  & \textbf{Palmer drought} \\ 
 & \textbf{} & \textbf{binary index} & \textbf{index}  & \textbf{severity index} \\ 
\hline\noalign{\smallskip} 
 Factual & 2021 & 2021 & 2021 & 2021 \\
 \hline\noalign{\smallskip}

Counterfactual & 1950 & 2021 & 2021 & 2021 \\
\noalign{\smallskip}\hline
\multicolumn{5}{c}{ }
\end{tabular}
}
\end{center}
\label{tab:statCFs}
\end{table}

Exploring the counterfactual scenario in this way allows us to identify the individual effect of GHG forcing 
under consistent ``background'' conditions as specified by ELI, PDSI and urbanization. We refer to the counterfactual scenario as ``statistical'' since it leverages the underlying statistical model to predict the distribution of temperature extremes for a setting of climate variables that did not actually occur (i.e., 1950 levels of GHG forcing and present-day ELI, PDSI, and urbanization). Finally, in a Bayesian framework we obtain posterior samples from all unknown quantities in Eq.~\eqref{eqn:gev_fam} and \eqref{eqn:gev_coefs} which allows us to propagate all statistical uncertainty to yield posterior distributions of the GEV upper bounds and risk probabilities. 

In addition to summarizing the inferred GEV upper bounds and risk probabilities for each scenario, we furthermore use the risk probabilities to construct an attribution statement for the anthropogenic drivers. Specifically, we compare the factual probability of the 2021 heatwave event (as specified by the 2021 TXx shown in Figure~\ref{subfig:maxTmax2021}) with corresponding probabilities from the counterfactual scenario, commonly referred to as probabilistic event attribution \citep{national2016attribution}. One way to summarize the anthropogenic influence on an event of interest is via the so-called ``risk ratio'' \citep{Paciorek2018}, i.e., the ratio of risk probabilities. The risk ratio is thus a unitless quantity that summarizes the multiplicative change in probability of a fixed threshold event due to anthropogenic influence; we calculate risk ratio via
\[
RR(\bs) = \frac{p_F(\bs)}{p_{C}(\bs)}
\]
Here, $p_F$ represents the risk probability for the factual scenario and $p_{C}$ represents the risk probability for counterfactual; see Eq.~\eqref{eqn:riskProb} for the definition of risk probability. $RR(\bs)$ attributes the isolated effect of anthropogenic GHG forcing on the probability of the heatwave occuring at location $\bs$; a risk ratio of greater than one indicates that  anthropogenic influence caused the 2021 PNW heatwave to become more likely, while a risk ratio of less than one indicates that anthropogenic influence caused the event to become less likely. We summarize best estimates and uncertainties in the risk ratios via their posterior distributions, which are obtained from the posterior distributions of the risk probabilities. 

\subsection{Alternative statistical models}\label{subsec:altMods}

Recall that the main objective of this paper is to determine whether a more appropriate statistical model (i.e., one that accounts for all known relationships in the data) can characterize the rarity of the 2021 PNW heatwave in an out-of-sample sense and hence enable meaningful conclusions regarding the anthropogenic influence on the probability of the event. However, the statistical model proposed in Section~\ref{sec:eva} represents three improvements upon the methodology used in, e.g., \cite{Philip21,emily2022dynamics}: (1) inclusion of additional spatio-temporal covariates (characterizing the influence of ELI, PSDI, and urbanization in addition to GHG forcing); (2) accounting for climatological dependence by including wet/dry regimes and orographic properties; and (3) accounting for spatial dependence in the tail of copula. In order to explicitly assess which aspect(s), if any, of our proposed model address the challenges of estimating out-of-sample probabilities related to an extremely rare heatwave, we consider three additional statistical models that can be viewed as special cases of the more general approach outlined in Section~\ref{sec:eva}:
\begin{itemize}
    \item[M1] Pointwise, GHG only: this represents the approach used to analyze in situ records in \cite{emily2022dynamics}. Marginally, we simplify Eq.~\eqref{eqn:gev_coefs} such that the location parameter is a linear function of GHG forcing and the scale and shape parameters are time-invariant. Then, GEV estimates are obtained independently for each weather station record (hence the term ``pointwise'') such that both climatological and data-level dependence are ignored.
    \item[M2] Pointwise, all covariates: this approach generalizes M1 by using the full suite of covariates (in both location and scale) described by Eq.~\eqref{eqn:gev_coefs}; however, like M1, both climatological and data-level dependence are ignored.
    \item[M3] Climatological dependence only, all covariates: this approach generalizes M2 by again using the full suite of covariates (in both location and scale) described by Eq.~\eqref{eqn:gev_coefs}; however, we now account for climatological dependence \textbf{only} and ignore data-level dependence. This is equivalent to the GEV-GP model in \citet{cooley2007bayesian}.
    \item[M4] All spatial, all covariates: this approach denotes the full model described in Section~\ref{sec:eva} that  includes all covariates and accounts for both climatological and data-level dependence.
\end{itemize}
These models are furthermore summarized in Table \ref{tab:statModels}. Fitting each of these models in a Bayesian framework allows us to calculate best estimates and uncertainties for 2021 GEV upper bounds, risk probabilities, and risk ratios. 

\begin{table}[!t]
\caption{Summary of the four extreme value analysis statistical models applied to the JJA seasonal maxima from the homogenized GHCN records over 1950-2020.}
\label{table:4_model_summary}
\begin{center}
{\small 
\begin{tabular}{clcccc}
\hline\noalign{\smallskip}
\multicolumn{2}{c}{\textbf{}} & \textbf{Include} & \textbf{Include ELI/} & \textbf{Account for}  & \textbf{Account for} \\ 
& & \textbf{GHGs?} & \textbf{PDSI/Urban?} &\textbf{clim. dep.?}  & \textbf{data-level dep.?} \\ 
\hline\noalign{\smallskip} 
M1 & Pointwise, GHG only & $\checkmark$ & & & \\
M2 &  Pointwise, all covts.&$\checkmark$ & $\checkmark$ & & \\
M3 & Clim. dep., all covts. &$\checkmark$ & $\checkmark$ & $\checkmark$ & \\
M4 & All spatial, all covts.&$\checkmark$ & $\checkmark$ & $\checkmark$ & $\checkmark$ \\
\noalign{\smallskip}\hline
\multicolumn{5}{c}{ }
\end{tabular}
}
\end{center}
\label{tab:statModels}
\end{table}

\section{Results} \label{sec:results}

\subsection{GEV coefficients}
\begin{figure}
    \centering
   \includegraphics[width=1\linewidth]{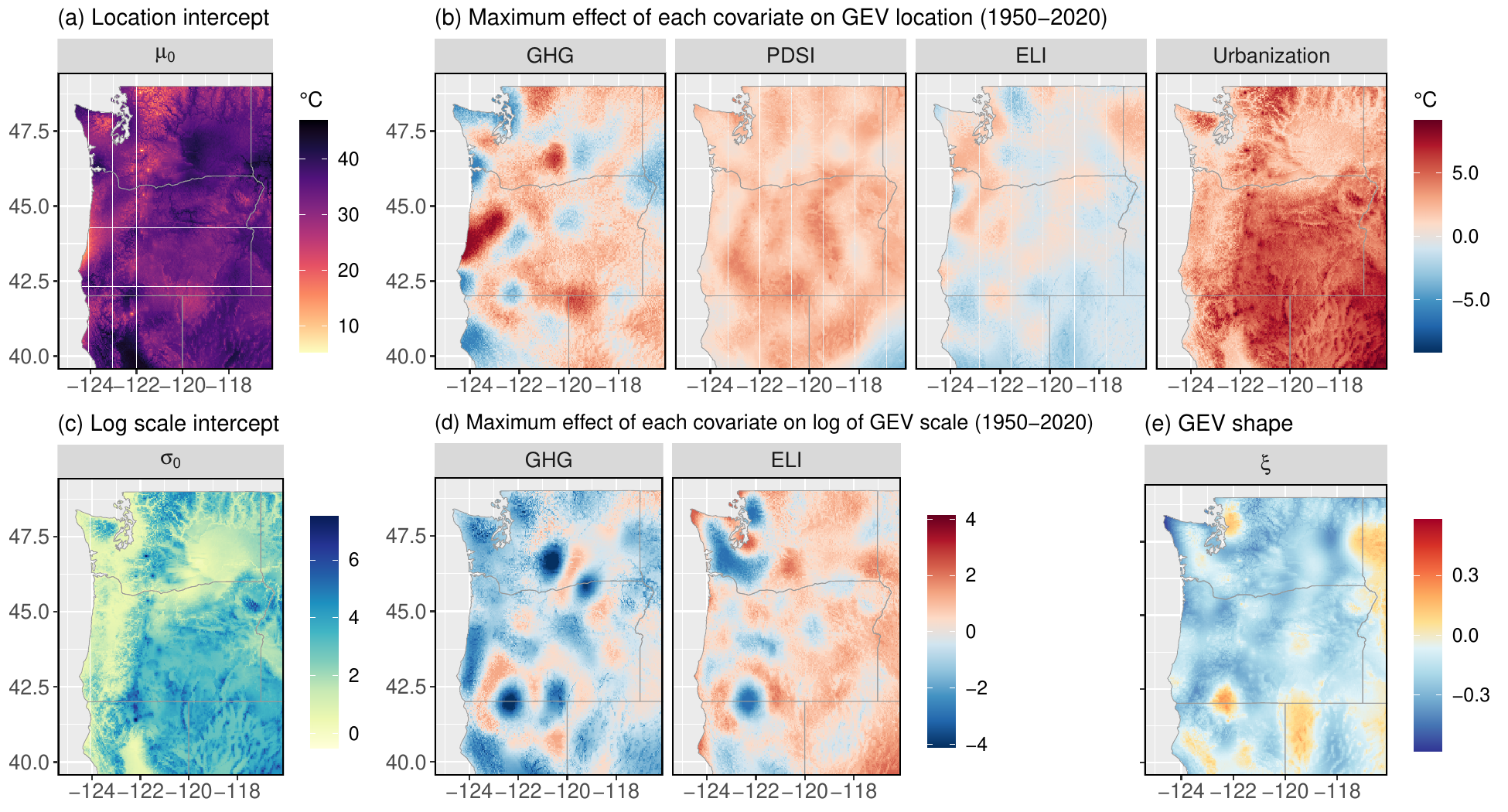}
    \caption{Posterior means of the maximum effects of GEV coeeficients $\{\mu_0(\bs),\ldots,  \mu_1(\bs),\sigma_0(\bs), \ldots, \sigma_2(\bs), \xi(\bs)\}$ under M4; see Table~\ref{table:4_model_summary} for the definition of M4 and see Eq.~\eqref{eqn:max_effect_GEV_coefs} to learn how to calculate the maximum effects.  }
    \label{fig:posterior_mean}
\end{figure}

From each MCMC update of basis coeffcients $\{\beta^{\gamma}_0,\beta^{\gamma}_1,\ldots, \beta^{\gamma}_B\} $, $\gamma\in \{\mu_0,\ldots,\mu_4,\sigma_0,\ldots, \sigma_2,\xi\}$, we plug each quantity into Eq.~\eqref{eq:TPS} and generate the GEV coefficient values $\{\gamma(\bs)\}$ on an unobserved $\sim 4$km $\times \sim4$km ($1/16^\text{th}$ degree) grid over the PNW region. Here we use the thin plate splines and topographical bases values calculated on the grid \textit{a priori}, and we update $\{\gamma(\bs)\}$ for each MCMC iteration. Figure \ref{fig:posterior_mean} displays the pointwise posterior medians of GEV coefficients (denoted by $\{\tilde{\gamma}(\bs)\}$). In order to directly compare the effect of each covariate, 
we multiply $\tilde{\mu}_j(\bs)$ and $\tilde{\sigma}_j(\bs)$ by the range of the time series over 1950-2021 (i.e., converting units from $^\circ$C per unit increase in covariate to just $^\circ$C). First, for $\tilde{\mu}_1(\bs)$ we show  
\begin{equation}\label{eqn:max_effect_GEV_coefs}
    \tilde{\mu}_1(\bs)\times\left[\max_{t}\{\text{GHG}_t\}-\min_{t}\{\text{GHG}_t\}\right].
\end{equation}
Figure \ref{fig:posterior_mean}(b) shows that an increase in GHG forcing led to a slight decrease of temperature near the Elliot Bay of Seattle and along the Pacific coast. However, an increase of at least 1$^\circ$C is observed in the majority of the PNW region, while the increase is more evident in the Yakima Valley of Washington ($\sim 4^\circ$C) and the Willamette Valley of Oregon($\sim 8^\circ$C). This in and of itself could be an attribution statement even if a few observed temperatures are still out of bounds (see later discussion in Section \ref{subsec:upperbounds}), since the human influence on extreme temperatures is insensitive to rarity \citep[see Figure 6 in][]{wehner2018early}. 

To show the maximum effect of PDSI, we examine the maximum effect on a pointwise basis and multiply $\tilde{\mu}_2(\bs)$ by the range of the time series $\{\text{PDSI}_t(\bs): t=1950,\ldots, 2021\}$ at each location $\bs$, i.e.,  
\begin{equation*}\label{eqn:max_effect_GEV_coefs_pdsi}
    \tilde{\mu}_2(\bs)\times\left[\max_{t}\{\text{PDSI}_t(\bs)\}-\min_{t}\{\text{PDSI}_t(\bs)\}\right].
\end{equation*}  
Recall that PDSI $>0$ indicates dry conditions, and PDSI $<0$ indicates wet conditions. From Figure \ref{fig:posterior_mean}(b), PDSI has a consistently positive effect on the GEV location, driving a  $\approx 1^\circ$C increase uniformly across the domain except for the lower right corner where there are no GHCN stations included in the analysis. 

Similarly to Eq.~\eqref{eqn:max_effect_GEV_coefs}, we multiply $\mu_3(\bs)$ and $\sigma_2(\bs)$ by the range of $\{\text{ELI}_t: t=1950,\ldots, 2021\}$. The effect of ELI on the GEV location is generally weaker than GHG and PSDI, and (like GHG forcing) its effect on typical extreme temperatures is varied over the PNW. The effect of ELI on the GEV scale is also heterogeneous, with an evident decrease in the Shasta Valley of northern California and west of the Cascade Mountains in Washington.

For the urban heat island effect, we multiply $\mu_2(\bs)$ by $1$ to assess the potential temperature increase brought by urbanization. If $\text{Urban}_t(\bs)\equiv 0$ for all $t$ at location $\bs$, the urban heat island effect is non-existent even though $\mu_4(\bs)$ might be positive in Figure \ref{fig:posterior_mean}; if $\text{Urban}_t(\bs)\equiv 1$ for all $t$, there is some urbanization effect but it stays constant over 1950-2021. It is more interesting to look into locations where $\text{Urban}_t(\bs)$ starts out as $0$ but changes to $1$ before 2020. For example, there are large areas near the Portland and Seattle metropolitan regions that were urbanized during 1950--2020; see Supplemental Figure \ref{fig:urban_mask}.  

\subsection{2021 upper bounds} \label{subsec:upperbounds}

\begin{figure}
    \centering
    \includegraphics[width=1.1\linewidth]{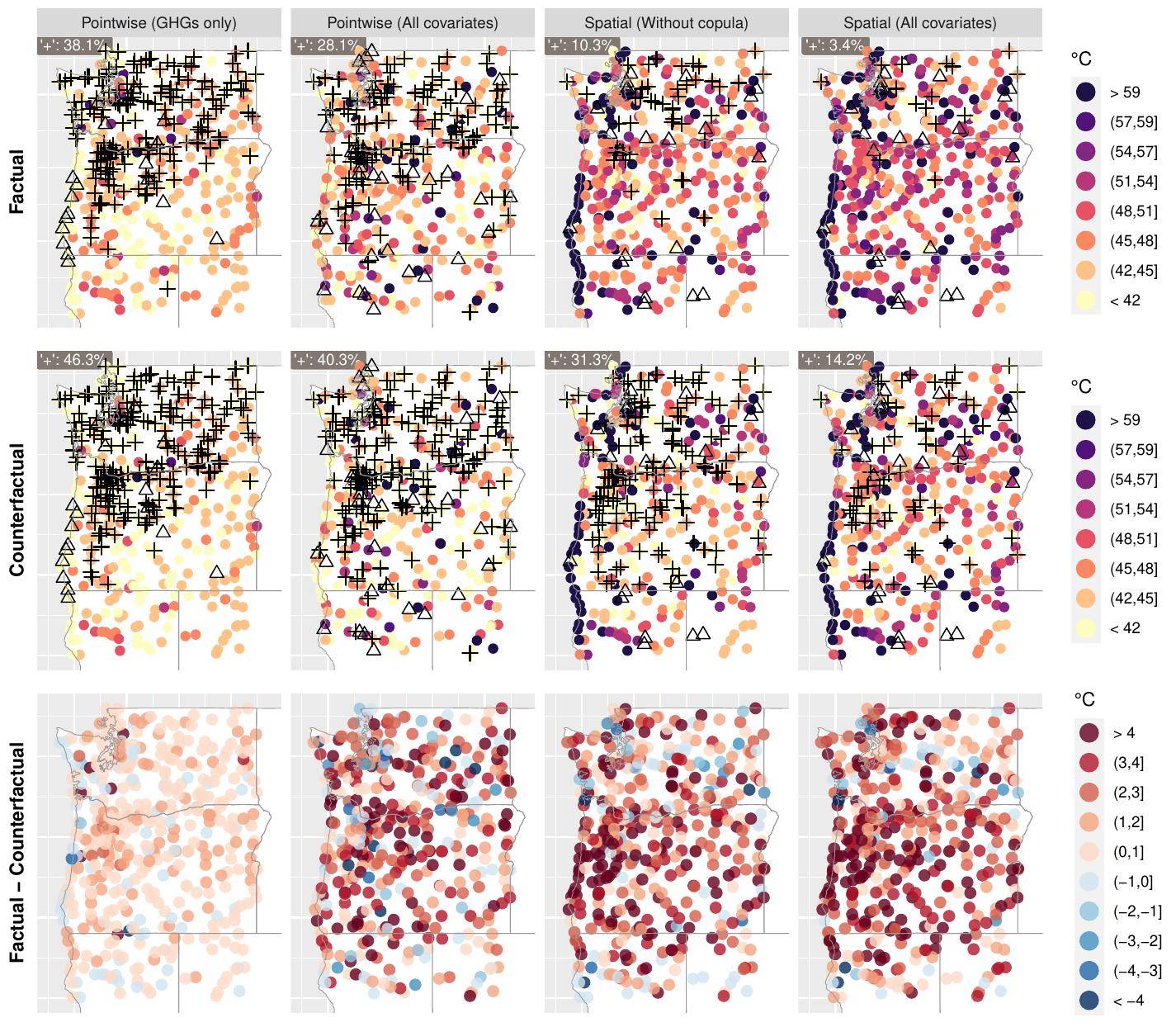}
    \caption{The posterior median of the GEV upper bound 
    under four statistical models (M1-M4; see Table~\ref{tab:statModels}). The first and second rows show upper bound estimates under the factual and counterfactual scenarios, respectively (see Table~\ref{tab:statCFs} for details). 
    For these rows, the `$\triangle$' signifies an infinite upper bound (corresponding to $\xi>0$), and `+' signifies stations for which the observed 2021 TXx exceeded the posterior median of the upper bound. The inset text in each panel displays the fraction of stations for which the upper bound is exceeded. The third row displays the differences in the upper bounds for each scenario.}
    \label{fig:GEV_upper_bounds}
\end{figure}

In Figure \ref{fig:GEV_upper_bounds}, we calculate the GEV distributional upper bounds using Eq.~\eqref{eqn:upperBound} and the out-of-sample MCMC updates of the marginal parameters under the four modeling settings outlined in Table \ref{tab:statModels}. The top row presents the pointwise medians of the posterior upper bounds in year 2021 using the factual conditions (see Table \ref{tab:statCFs}), in which the pointwise analyses (M1 and M2) produced relatively low upper bounds such that over 38\% and 28\%, respectively, of the stations had 2021 TXx values that exceed their corresponding theoretical upper bounds. In comparison, accounting for spatial dependence, even only at the climatological level, greatly improved the predictability of the 2021 heatwave event while increasing the GEV upper bounds consistently across PNW region: under M3, only 10.3\% of stations have 2021 TXx that exceed the posterior median upper bound. Interestingly, the GEV shape parameters at many coastal stations changed from being positive in M1 to being negative but close to zero in M3 and M4 such that the upper bounds are finite but very large. In particular, the cluster of inexplicable 2021 TXx values near the Portland and Seattle-Tacoma metropolitan area in M1-M3 disappeared in M4 due to the incorporation of the urbanization index and extremal copula modeling. Further accounting for extremal dependence in M4 reduced the number of unexplainable events to just 3.4\% of stations, demonstrating that accounting for both climatological and extremal dependence are critical for making the PNW heatwave event more predictable in an out of sample sense. Additionally, in one of our previous analyses in which we basically followed the M4 settings but excluded the urban binary index from the model, the proportion of stations with a 2021 TXx value exceeding the posterior median of upper bound was 9.4\%, compared to 3.4\% under M4 with the urban binary index in upper right panel of Figure \ref{fig:GEV_upper_bounds}. Nonetheless, there are still some observations that exceed the posterior median GEV upper bound in spite of using a more sophisticated statistical model with the urbanization index, albeit the upper limits of our 95\% credible intervals of the upper bounds from M4 contain the 2021 TXx at all stations. 

The middle and bottom rows of Figure \ref{fig:GEV_upper_bounds} examine the isolated effect of GHG forcing under the counterfactual scenario, which show that many more stations in Washington and Oregon (between 8-12\% more) would have failed to contain the 2021 TXx values if the GHG stayed at the 1950 level. Especially under M3 and M4, differences in the posterior medians of the upper bounds exceeds 3$^\circ$C in majority of the stations in Oregon; i.e., increases in GHG forcing drive increases in GEV upper bounds by more than 3$^\circ$C. Although M2 under the factual conditions failed to contain a large fraction of the 2021 TXx measurements, the differences in upper bounds are much higher than M1 and even comparable to spatial analyses. This indicates that including more covariates will augment the anthropogenic effects and increase the factual upper bounds but not enough to contain significantly more 2021 TXx values compared to M1. Interestingly, under all modeling settings and factual/counterfactual conditions, there are practically no stations in northern California and eastern Oregon whose 2021 TXx values exceeded the upper bounds, indicating a not-so-extreme heatwave in that region.

\subsection{Attribution: counterfactual risk probabilities and risk ratios}\label{subsec:attribution}

    
    


\begin{figure}[!ht]
\centering
    
    \includegraphics[width=\linewidth]{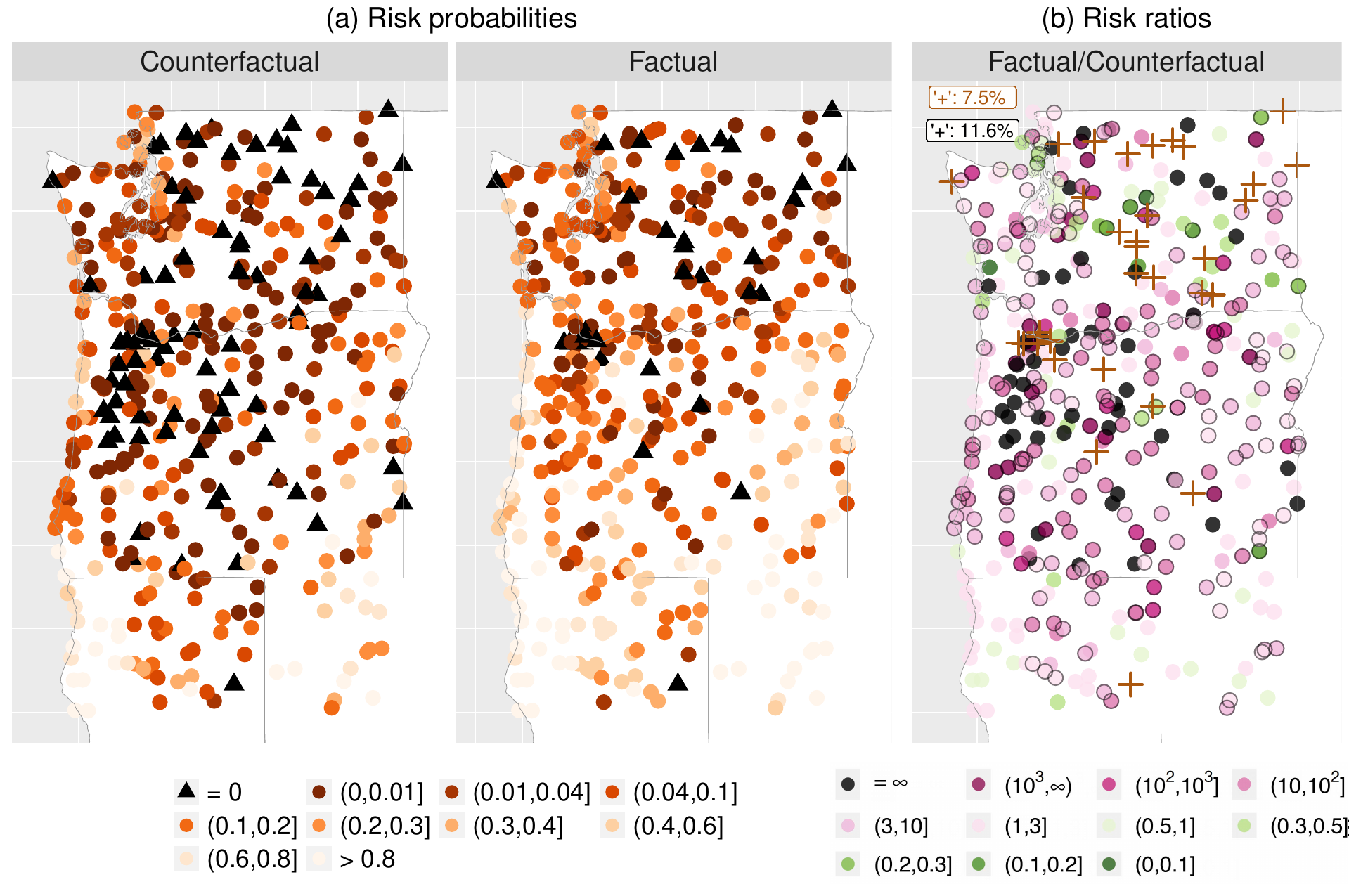}
    \vskip -1em
    \caption{Subfigure 
    (a) shows the station-specific best estimates (posterior medians) of the risk probabilities calculated from statistical model M4 for the Counterfactual (left) and Factual (right) climate scenarios, and subfigure 
    (b) shows their ratio. For the risk probabilities, solid black triangles indicate gauged locations for which the risk probability best estimate is zero; for the risk ratios, solid black circles denote $RR(\bs) = \infty$ (wherein the counterfactual risk probability is zero but the factual risk probability is nonzero) and yellow `$+$' shows where the risk ratios are undefined (i.e., both counterfactual and factual risk probabilities are zero). In the rightmost panel, points that are plotted with additional `$\circ$' sign indicates that risk ratio estimates are statistically significantly different from $1$.
    }
    \label{fig:Exceed_2021_probs}
\end{figure}

In order to assess whether or not anthropogenic factors had a meaningful influence on the probability of the 2021 heatwave in a Granger sense \citep{granger1969investigating, risser2017attributable}, we calculate both location-specific estimates of the risk probabilities and risk ratios for the PNW region (defined by the 40$^\circ$N–49$^\circ$N, 125$^\circ$W–117$^\circ$W bounding box); see Figure~\ref{fig:Exceed_2021_probs}. For attribution results we now focus in on the estimates obtained from the model described in Section~\ref{sec:eva} (labeled M4 in Table~\ref{tab:statModels}) since this approach best accounts for known relationships in the data and minimizes the fraction of stations for which the GEV upper bounds are exceeded by the 2021 extreme temperatures.

First, consider the posterior median estimates of the risk probabilities for the factual $p_F(\cdot)$ and counterfactual $p_C(\cdot)$ scenarios, shown in Figure~\ref{fig:Exceed_2021_probs}(a). Recall that $p_F(\cdot)$ and $p_C(\cdot)$ quantify the probability of experiencing an extreme daily maximum temperature measurement at least as large as what was observed during the heatwave under each climate scenario; furthermore, these probabilities represent the best seasonal forecast for extreme daily temperatures in advance of the 2021 summer using a statistical model given the values of the covariates in Eq.~\eqref{eqn:gev_coefs}. The risk probabilities also summarize how unusual the heatwave event was at each station. The counterfactual risk probabilities are generally quite small: for a large majority of stations, the event had a probability of less than 0.1 and in many cases less than 0.01. For the factual scenario, risk probability estimates are also small, but not as small as for the counterfactual scenario: factual risk probabilities are generally between 0.05 and 0.3. As would be expected from the discussion around upper bounds in Section~\ref{subsec:upperbounds}, there are cases where the posterior medians of the risk probability are zero (denoted by black triangles in Figure~\ref{fig:Exceed_2021_probs}), meaning that the 2021 heatwave temperatures would have been considered impossible in the counterfactual or factual scenario. 
However, there are many stations for which the 2021 temperatures were quite ``normal,'' even in the counterfactual scenario: across Nevada and much of northwest California, the observed heatwave temperatures had more than a 0.5 probability. 
In the meantime, there are still a small number of stations for which the factual risk probability estimates are exactly zero -- in other words, for several stations we would have concluded that the 2021 temperatures were impossible even with increases to GHG forcing. However, it is the case that $p_F(\bs) = 0$ if and only if $p_C(\bs) = 0$, i.e., if the event was considered impossible in the factual scenario than it would also have been considered impossible in the counterfactual scenario. 


Next, consider the posterior median estimates of the risk ratios (i.e., the factual risk probability divided by the counterfactual risk probability), shown in Figure~\ref{fig:Exceed_2021_probs}(b). Across the PNW, the dominant color in Figure~\ref{fig:Exceed_2021_probs}(b) is purple/magenta (i.e., $RR(\bs)>1$), indicating that the 2021 heatwave event was made more likely by increases to GHG forcing. The risk ratios are often larger than 3, indicating that increases to GHG forcing resulted in a 3-fold increase in the probability of the extreme temperatures associated with the heatwave (and often much more than a 3-fold increase). In many cases, these increases in risk probability are statistically significant, in the sense that the lower bound of the 95\% credible interval for the risk ratio is also greater than 1 (indicated by black circles around points in Figure~\ref{fig:Exceed_2021_probs}(b)). 
This is a powerful result: using a more refined statistical model, we can  both quantify nonzero probabilities for the PNW heatwave event as well as determine statistically significant changes to this probability \textit{in an out-of-sample sense} -- this was previously impossible \citep[again, see][]{Philip21,emily2022dynamics}.
The presence of zeros in the best estimates of the counterfactual risk probability means that in many cases the risk ratio is infinite; however, there are a still $\sim 7$\% of stations for which the risk ratio is undefined (i.e., both risk probabilities are zero, such that $RR = 0/0$), even using our best statistical model. (Note that the 7.5\% of stations with undefined risk ratios differs from the 3.4\% of stations for which the factual GEV upper bounds were exceeded due to the nonlinearity of posterior calculations.)
Notably, there are some cases where the risk ratio is less than one (very few stations significantly so). The locations of these stations coincides with the regions where the GHG coefficient is negative; see Figure \ref{fig:posterior_mean}(b). They also correspond to the stations that have very low risk probabilities under both factual and counterfactual scenarios.

In summary, we now have a solution to the original problem posed by the severity of the heatwave event for existing attribution methods: while na\"ive statistical methods fail to quantify nonzero probabilites of the event (with or without climate change) and hence cannot make confident attribution statements, by using a sophisticated statistical model that more appropriately accounts for known features of the data (i.e., covariates, climatological dependence, and extremal dependence in the copula) we are able to robustly quantify the extent to which anthropogenic GHG emissions increased the likelihood of the observed extreme temperatures in 2021.

\section{Conclusion} \label{sec:conc}
In this paper, we built a novel statistical model that accounts for the spatial coherence of both the climatological coefficients while also considering the natural and anthropogenic drivers for extreme temperatures. Using a Bayesian hierarchical framework, our method improves on the results from both \citet{Philip21} and \citet{emily2022dynamics} as more than 38\% of the in situ 2021 TXx values could not be anticipated in their out-of-sample analyses, compared to 3.4\% from our best model fit. As a result, we were able to estimate risk probabilities and make attribution statements for most GHCN stations while also quantifying the uncertainties for the estimates of the risk ratios. We showed that the increases in GHG forcing resulted in at least 3-fold increase in the probability of the extreme temperatures associated with the heatwave at the majority of the PNW locations in the US. Furthermore, we performed analyses under a set of simpler models to isolate the contributions from each component of the model to the improvements over predicting the 2021 PNW heatwave. We found that borrowing strength from neighboring stations, even only at the climatological level, maximizes the information contained in the historical data which sheds more light on the extreme events.

It should be noted here that while our analyses focus on the PNW region in the US, the most intense part of the heatwave occurred in British Columbia because the associated heat dome was centered there. However, we confined our analyses to the PNW region within the United States due to a lack of the homogenized temperature data in Canada from GHCN. \citet{McKinnon_Simpson} increased the spatial coverage of British Columbia by including station data archived by \citet{canada_data} and the sub-daily measurements in the Integrated Surface Database \citep{smith2011integrated}. In future work we plan to include in situ observations from both Canada and the US 
and re-run the analysis to get a more complete picture of the heatwave and to evaluate the anthropogenic and natural contributions to the event for the larger spatial domain. 

\citet{emily2022dynamics} argued that the failure to contain the 2021 TXx values using the GEV upper bounds resulted from a 
violation of the ``IID'' assumption in the univariate extreme value theory, which requires the samples from each block (season in our case) to be statistically independent and identically distributed. As explained in the introduction, the underlying meteorological conditions responsible for the extreme temperatures are truly unique and different in 2021, indicating that the observed TXx from 2021 do not arise from the same distribution as the TXx values from the preceding 71 years. Including covariates to model year-to-year changes in the marginal GEV distribution relaxes this assumption such that we only require the ``residuals'' to be IID. Nonetheless, using covariates in the GEV parameters alone within a single station analysis framework is not enough to characterize the extreme 2021 temperatures (as shown in both \citealp{emily2022dynamics} and models M1 and M2; see Figure~\ref{fig:GEV_upper_bounds}). Instead, the real difference-maker is accounting for spatial dependence in extreme temperatures: even accounting for just climatological dependence in the GEV covariates drastically reduces the fraction of ``unexplainable'' 2021 temperatures (28.1\% to 10.3\% for the factual scenario). However, incorporating physical covariates \textit{and} accounting for climatological dependence in these quantities is also not enough: it is only when additionally accounting for data-level dependence via the copula that we can provide the best characterization of the 2021 extreme temperatures.

In Section \ref{sec:results}, we focused on posterior median summaries for determining exceedance of upper bounds and assessing the risk ratios between the factual and counterfactual scenarios. A principal benefit of using a Bayesian framework is that one can  summarize uncertainty robustly via either 95\% credible intervals or non-binary probabilities of upper bound exceedance. As noted previously, it is the case that the upper limits of 95\% credible intervals of the upper bounds for statistical model M4 contain the 2021 TXx at all stations! As such, it could be more informative to instead summarize the rarity of the 2021 TXx values with probabilities related to the upper bound failing to contain the event at each station using the posterior samples of our model parameters. However, for simplicity of presentation, in this manuscript we opted to show results for 
the upper bound 
posterior medians.

Finally, it is noteworthy that the flexible scale mixture model described in Section~\ref{sec:eva} concluded that TXx measurments in the PNW are asymptotically independent (AI; see Figure \ref{fig:chi_emp} in Appendix \ref{sec:chi_emp}). A similar property emerged for measurements of extreme precipitation \citep{zhang2022storm}; however, extreme precipitation events are generally more localized than extreme temperature events, such that we might expect the analysis in this paper to yield asymptotic \textit{dependence} (AD). As discussed in \citep{zhang2022storm}, one reason for extreme temperatures exhibiting AI is the size of the spatial domain considered in this analysis: analyzing smaller spatial domains may cause extreme temperatures to become asymptotically dependent. This sheds light on the importance of building a more flexible model that exhibits nonstationary tail dependence; we are currently pursuing such methodology in related work.


\addcontentsline{toc}{section}{References}
\bibliography{main}

\clearpage
\appendix
\section*{Supplementary material}
\numberwithin{equation}{section}
\numberwithin{figure}{section}
\numberwithin{table}{section}

\section{Supplementary figures}
\begin{figure}[!ht]
    \centering
     \begin{subfigure}[t]{.328\linewidth}
\centering\includegraphics[height=1.05\linewidth]{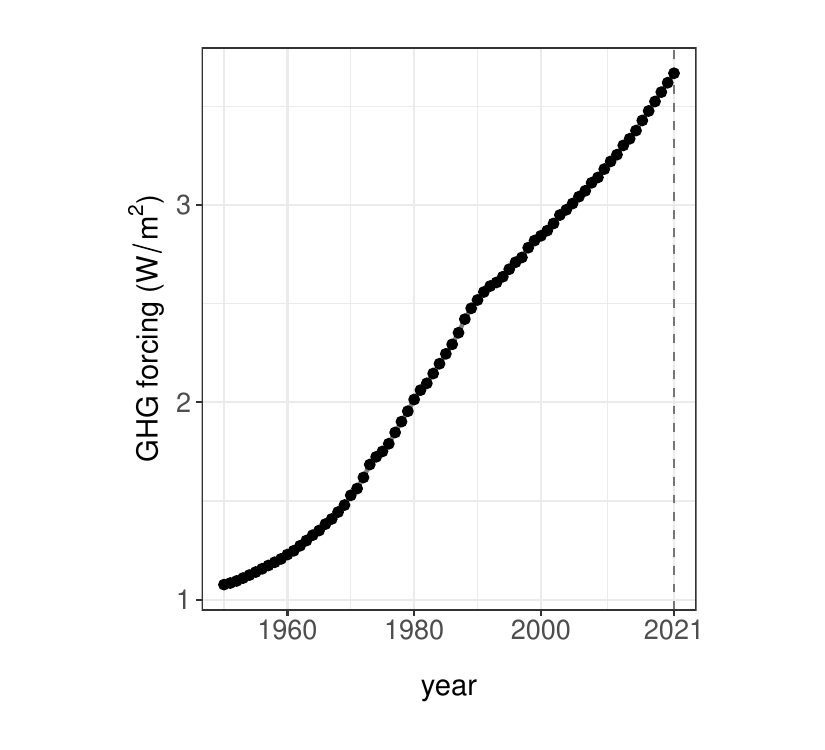}
\caption{Time series of GHG forcing}
    \label{subfig:GHG}
    \end{subfigure}
    \begin{subfigure}[t]{.328\linewidth}
\centering\includegraphics[height=1.05\linewidth]{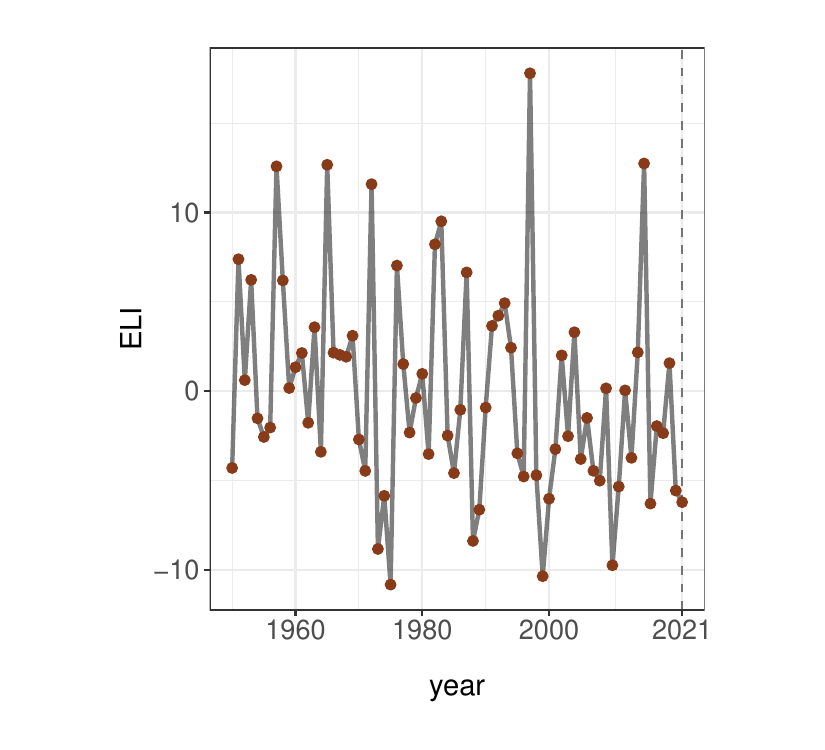}
\caption{Time series of average JJA ENSO Longitude Index (ELI)}
    \label{subfig:ELI}
    \end{subfigure}
    \begin{subfigure}[t]{.328\linewidth}
\centering\includegraphics[height=1.05\linewidth]{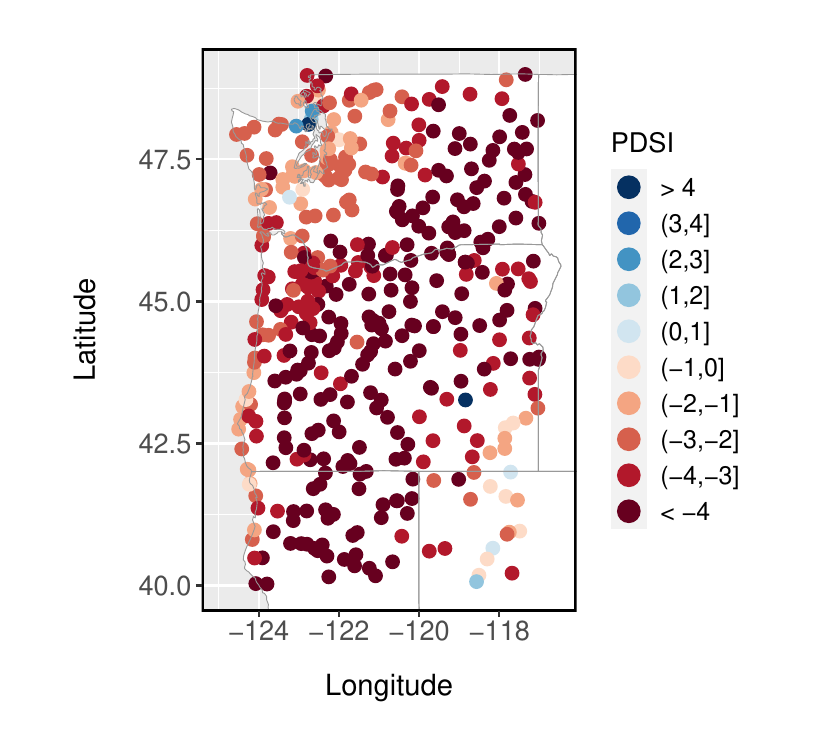}
    \caption{Average JJA PDSI of 2021}
    \label{subfig:pdsi}
    \end{subfigure}
    \caption{
    \subref{subfig:GHG} time series of radiative forcing due to greenhouse gas emissions, 
    \subref{subfig:ELI} the time series of average JJA ENSO Longitude index, 
    and \subref{subfig:pdsi} the average JJA Palmer Drought Severity Index (PDSI) in 2021.} 
    \label{fig:data_covariates1}
\end{figure}



\begin{figure}[!ht]
\centering
\includegraphics[width=\textwidth]{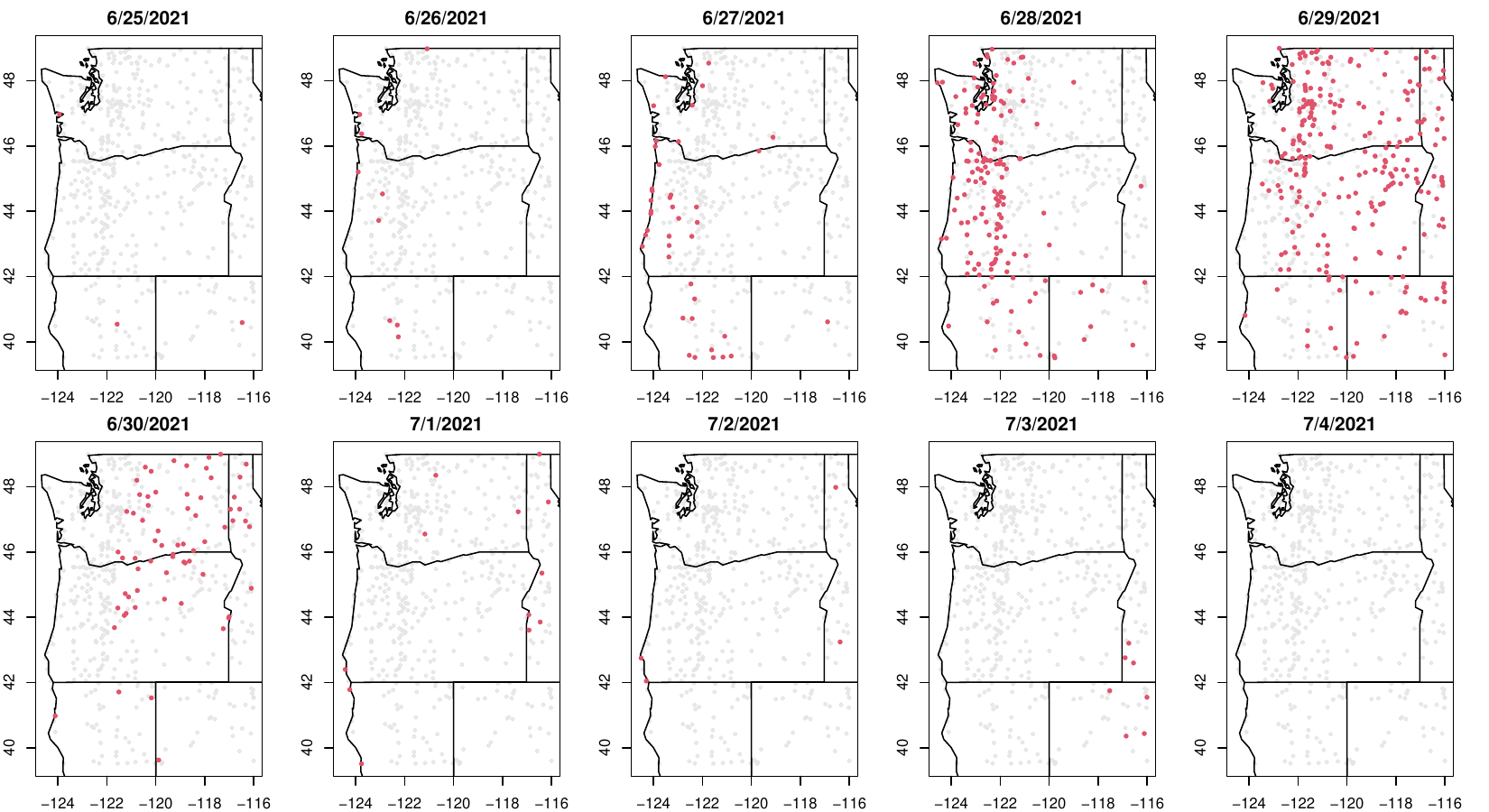}
\caption{Calendar date on which the maximum daily maximum temperature (TXx; $^\circ$C) occurred, from June 25, 2021 through July 4, 2021, for all gauged locations in the GHCN-D \citep{Menne2012} database that have at least one non-missing daily maximum temperature measurements over these ten days (538 total gauged locations). }
\label{fig:dateOfTXx}
\end{figure}

\begin{figure}[!ht]
\centering
\includegraphics[width=0.9\textwidth]{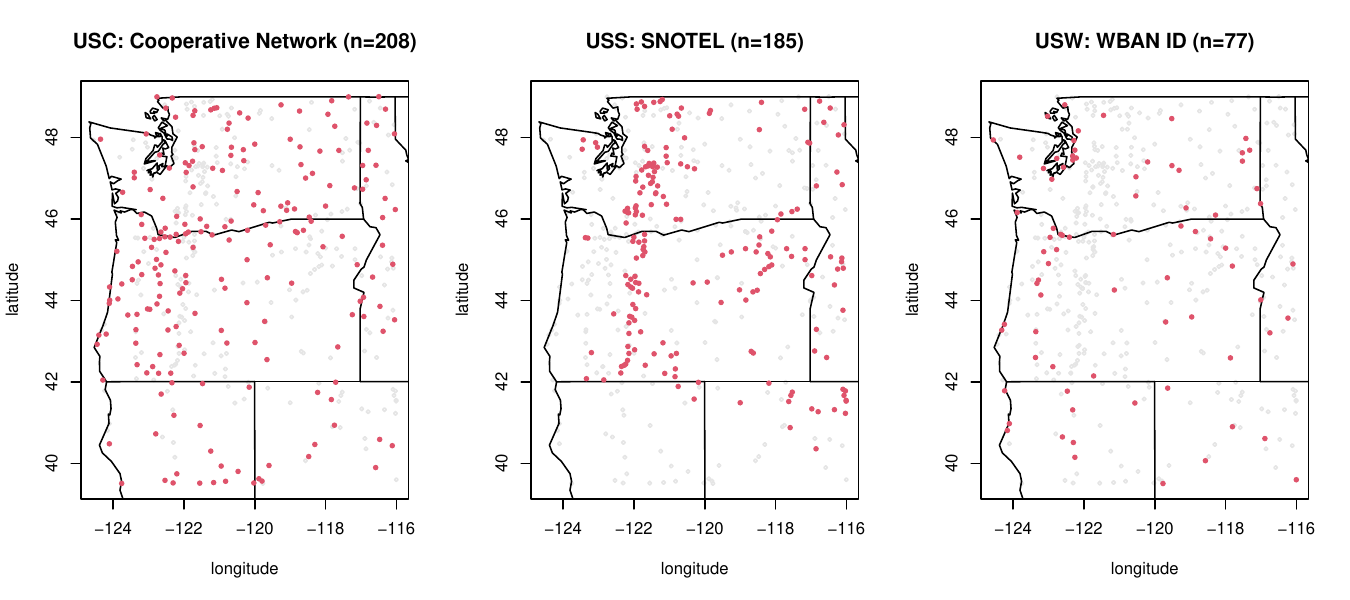}
\caption{Database code for each of the $n=470$ GHCN-D records that have six nonmissing daily maximum temperature measurements for June 26, 2021 through July 1, 2021.}
\label{fig:GHCNcode}
\end{figure}

\begin{figure}[!ht]
\centering
\includegraphics[width=\textwidth]{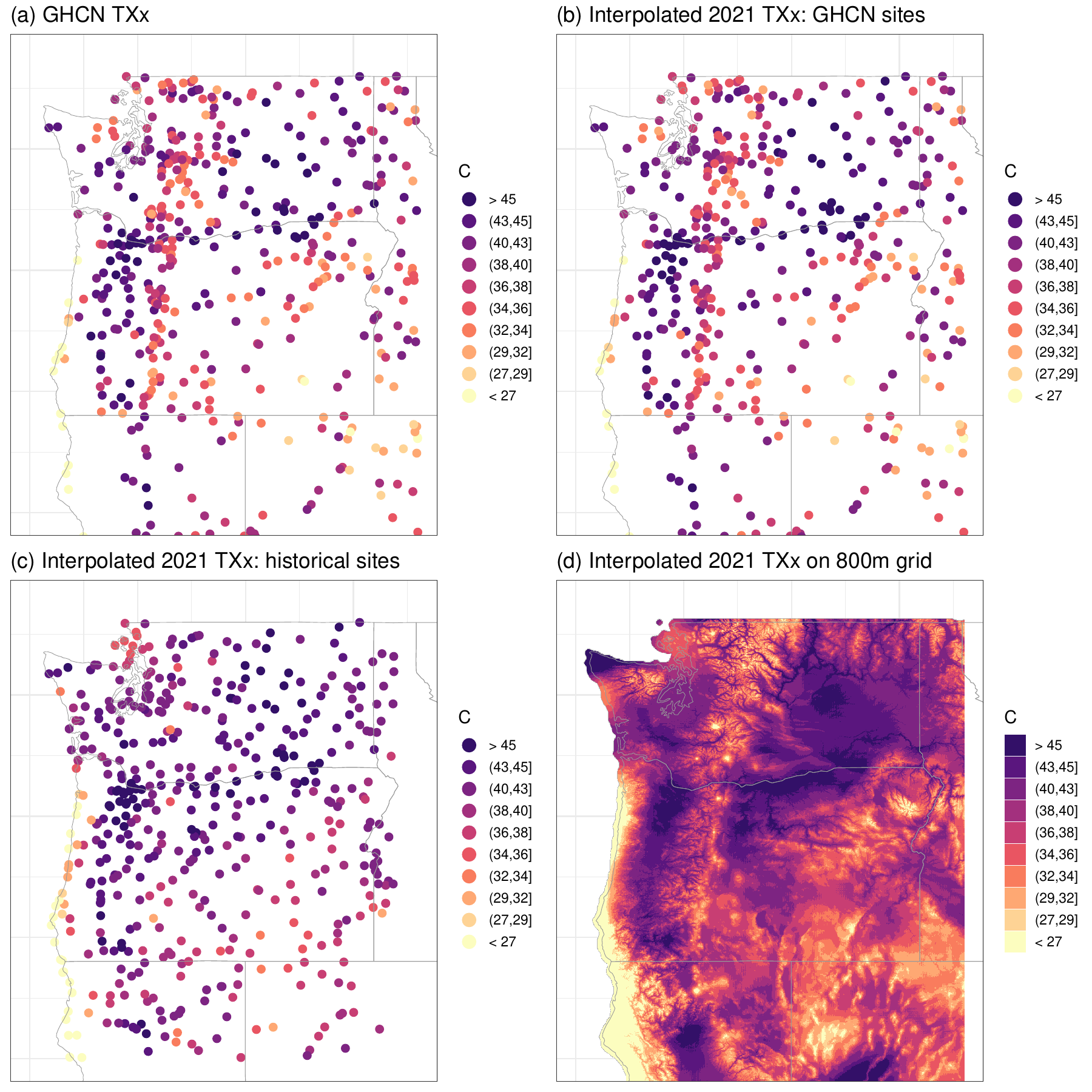}
\caption{Maximum daily maximum temperature (TXx; $^\circ$C) at the $n=470$ GHCN-D gauged locations that have six nonmissing daily maximum temperature measurements for 06/26/2021-07/01/2021 (panel a), with interpolated TXx measurements at the same sites using the spatial model defined in Section~\ref{sec:data} (panel b). Corresponding estimates for the $n=438$ gauged locations with homogenized records over 1950-2020 used for the historical trend analysis \citep{rennie2019development} are shown in panel (c), with interpolated estimates on a regular 800m grid over the PNW domain (panel d).}
\label{fig:pnw_TXx}
\end{figure}

\begin{figure}[!ht]
\centering
\includegraphics[width=\textwidth]{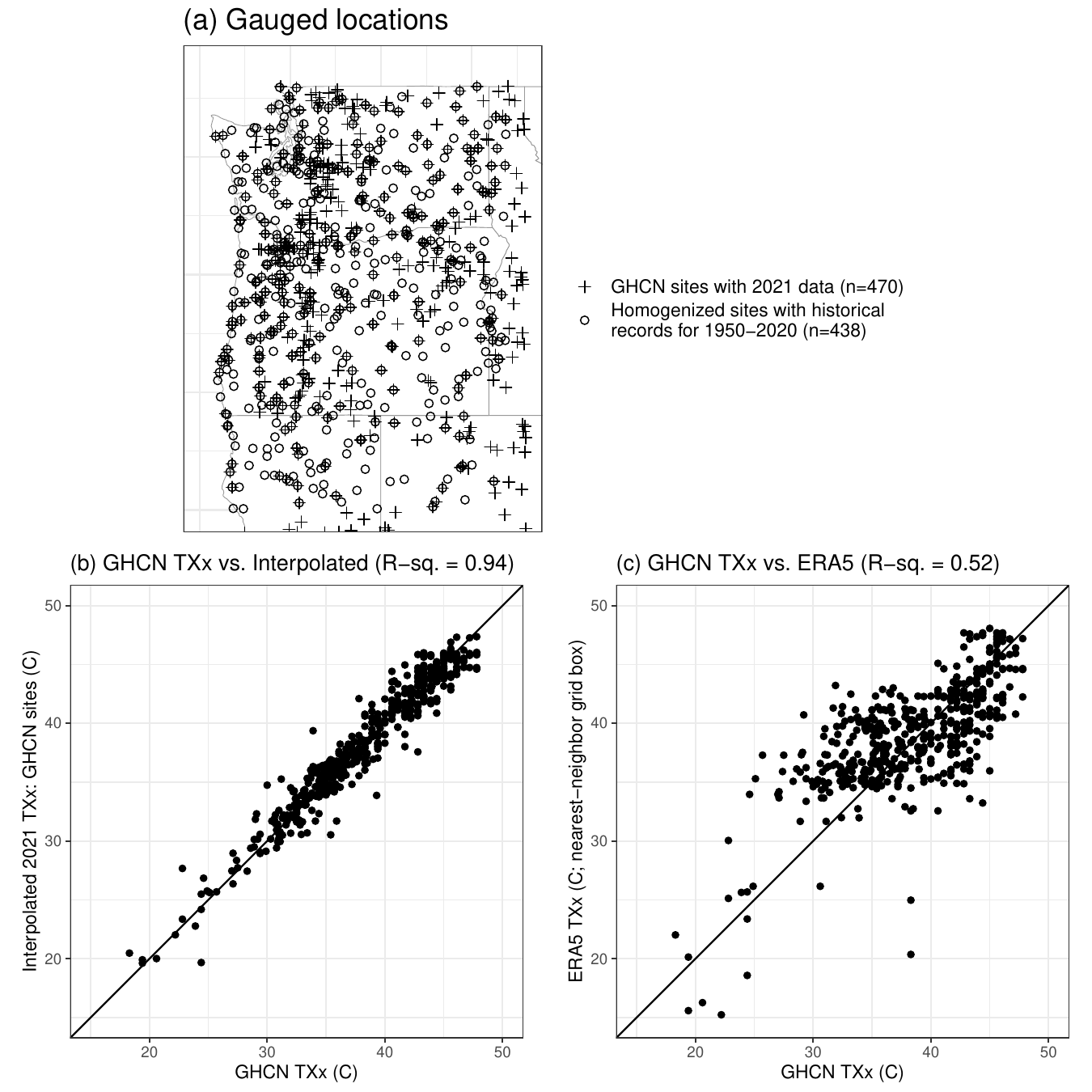}
\caption{Panel (a) shows the spatial distribution of the GHCN-D sites with 2021 daily maximum temperature measurements and the gauged locations with homogenized records for historical analysis.
Panel (b) compares the GHCN-D TXx measurements at the $n=470$ sites along with interpolated estimates at the same locations; for reference, panel (c) shows the GHCN-D values compared with nearest-neighbor ERA5 reanalysis grid boxes.
}
\label{fig:compareTXx}
\end{figure}

\begin{figure}[!ht]
\centering
\includegraphics[width=\textwidth]{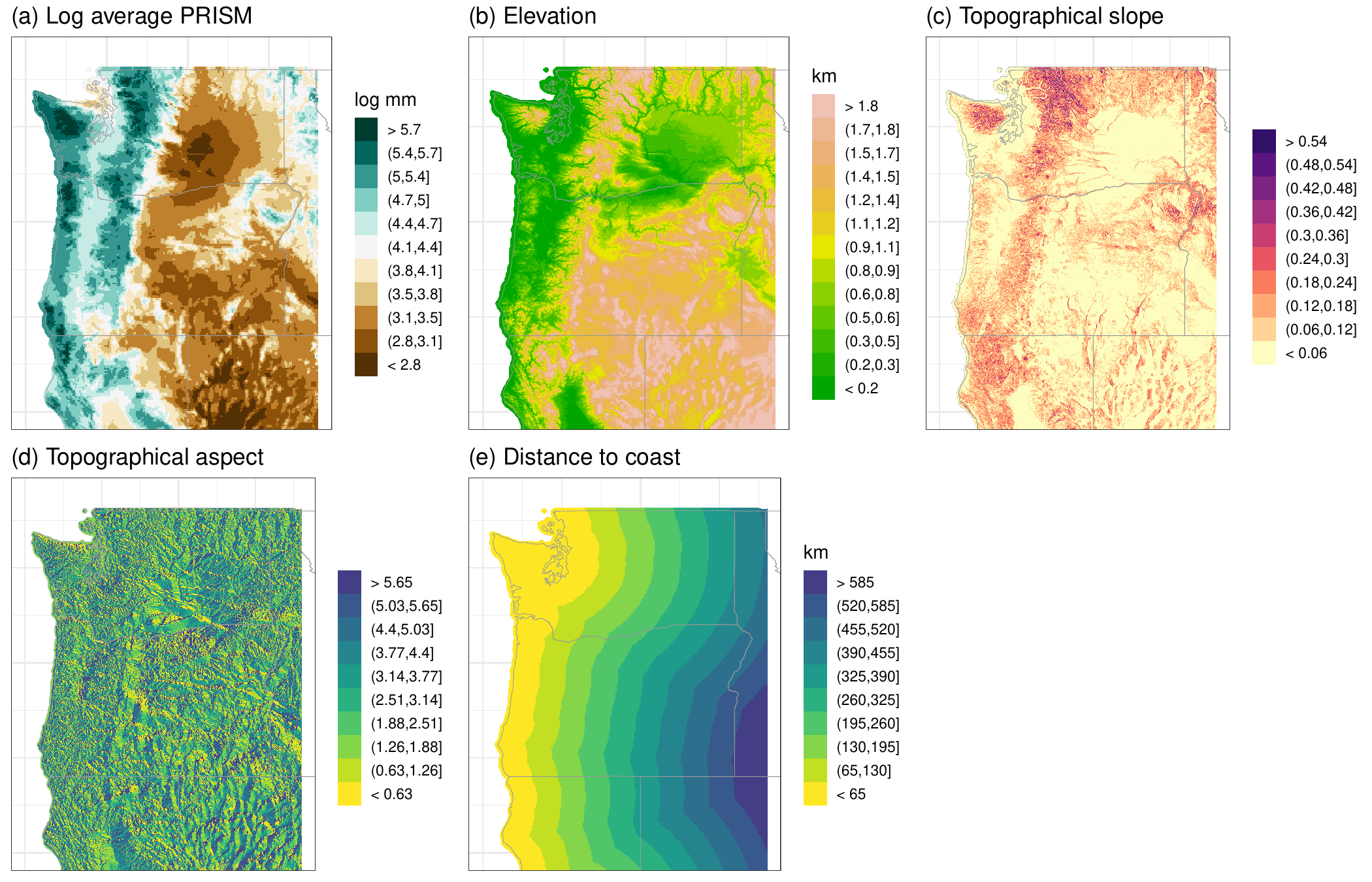}
\caption{Topographical covariates used to model spatial nonstationarity in the GEV parameters.}
\label{fig:topoCovt}
\end{figure}

\begin{figure}[!ht]
\centering
\includegraphics[width=0.6\textwidth]{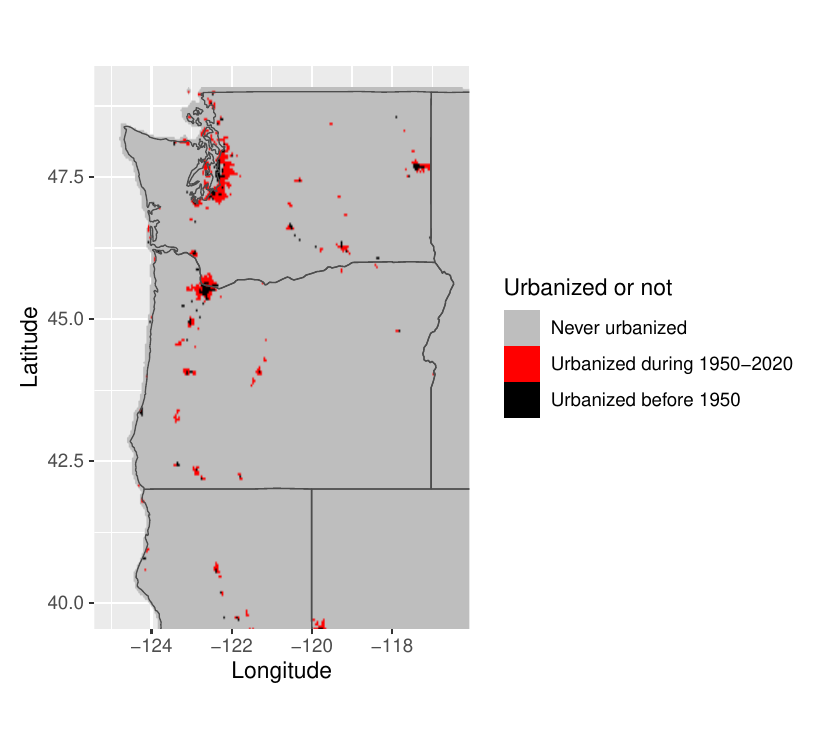}
\caption{Raster plot showing the locations that were urbanized during 1950-2020 (in red), and the locations that were already urbanized by 1950 (in black) or never urbanized until 2020 (in gray).}
\label{fig:urban_mask}
\end{figure}

\clearpage
\section{Hierarchical model}\label{sec:hierarchy_model}
Conditioning on $\gamma(\bs)$ and the latent Gaussian process $\boldsymbol{W}_t$, the likelihood for $\boldsymbol{Y}_t$ is independent between locations due to the presence of the white noise process $\{\epsilon(\bs)\}$ and can be described by
\begin{equation}\label{full_marginal}
\varphi\left(Y_{tj}\vert R_t, \boldsymbol{W}_t,\delta,\btheta_j,\tau^2\right)=
\frac{1}{\tau^2}\phi\left(\frac{X_{tj}- R_tW_{tj}}{\tau}\right)\frac{f_{Y\vert \btheta_j, t}(Y_{tj})}{f_{X\vert \delta,\tau^2}\left(X_{tj}\right)},
\end{equation}
in which $X_{tj}=F_{X\vert \delta,\tau^2}^{-1}\circ F_{Y\vert \btheta_j, t}(Y_{tj})$, $f_{Y\vert \btheta_j, t}$ and $f_{X\vert \delta,\tau^2}$ are the density functions of $F_{Y\vert \btheta_j, t}$ and $F_{X\vert \delta,\tau^2}$, and $\phi$ is the probability density function of the standard Normal distribution. The likelihoods for all locations from different time replicates are simply multiplied together. Following \citet{gelfand2012hierarchical}, we can specify the data model:
\begin{equation*}
\prod_{t=1950}^{2017}\prod_{j=1}^{\numStations}\varphi\left(Y_{tj}\vert R_t, \boldsymbol{W}_t,\delta,\btheta_j,\tau^2\right)=
\prod_{t=1950}^{2020}\prod_{j=1}^{\numStations}\frac{1}{\tau^2}\phi\left(\frac{X_{tj}- R_tW_{tj}}{\tau}\right)\frac{f_{Y\vert \btheta_j, t}(Y_{tj})}{f_{X\vert \delta,\tau^2}\left(X_{tj}\right)}.
\end{equation*}
It is worth noting that many papers in the spatial extremes literature, such as \citet{huser2017bridging}, \citet{huser2019modeling} and \citet{zhang2021hierarchical}, use empirical or approximate marginal transformations instead of modeling the marginal distributions formally like we do in Eq.~\eqref{eqn:marginalTrans}.

The distribution function $F_{X\vert \delta,\tau^2}$ and the density function $f_{X\vert \delta,\tau^2}$ are both convolutions on the distribution/density functions of the Gaussian scale mixture and the nugget. Therefore, they involve irregular integrals that do not have an analytical form; see the Appendix of \citet{zhang2021hierarchical}. We utilize the numerical integration functions from the Python library \texttt{scipy.integrate} to efficiently compute $F_{X\vert \delta,\tau^2}$ and $f_{X\vert \delta,\tau^2}$ while calling the \texttt{C++} defined integrands via modules \texttt{ctypes} and \texttt{scipy.LowLevelCallable}.

For the marginal GEV fields, the spline basis expansion can be written as
\begin{equation}\label{eqn:spline_basis_rep}
    (\gamma(\bs_{1}),\ldots,\gamma(\bs_{\numStations}))^T\vert\boldsymbol{\beta}^{\gamma}=  (\boldsymbol{1}_{\numStations}, \boldsymbol{S})\boldsymbol{\beta}^{\gamma},
\end{equation}
in which $\gamma\in\{\mu_0,\mu_1,\log \sigma,\xi\}$, $\boldsymbol{\beta}^{\gamma}=(\beta^{\gamma}_0,\beta^{\gamma}_1,\ldots, \beta^{\gamma}_B)^T$ and $\boldsymbol{S}$ is a $\numStations\times B$ matrix whose $j$th column is $(b_j(\bs_1),\ldots, b_j(\bs_{\numStations}))^T$. In the meantime,
\begin{align*}
    \beta^{\gamma}_i|\sigma^{\gamma}_{\beta}&\sim N(0,(\sigma^{\gamma}_{\beta})^2),\; i=0,1,\ldots,B.
\end{align*}

To further ease the computational burden and improve the mixing of the metropolis algorithm, we partition the index set $\{1, \ldots, B\}$ into $K$ equal-sized subsets.
For each MCMC draw, we perform block random-walk updates for the basis coefficients one subset at a time. More specifically, if $I_k$ is the index set of the $k$th subset ($k=1,\ldots, K$), denote the $m$th draw of the basis coefficients within the $k$th subset by $\boldsymbol{\beta}_k^{\gamma^{(m)}}=(\{\beta_i^{\gamma^{(m)}}:i\in I_k\})^T$. Iterating $k$ from $1$ to $K$, we take the current values of $\mu^{\gamma}$ and draw $\boldsymbol{\beta}_k^{\gamma^{(m)}}$ as follows:
\begin{enumerate}[(1)]
    \item Propose 
    \begin{equation*}
        \boldsymbol{\beta}_k^{\gamma*}\sim N\left(\boldsymbol{\beta}_k^{\gamma^{(m-1)}},\lambda_k\Sigma_{k,\gamma}\right),
    \end{equation*}
    in which $\lambda_k$ is a positive scaling constant which is automatically tuned every other 10 iterations using the adaptive metropolis algorithm proposed in \citet{shaby2010exploring} so that the optimum acceptance rate $0.234$ is achieved, and $\Sigma_{k,\gamma}$ is a positive definite proposal covariance matrix which is also updated every other 10 iterations so that it looks like the sample covariance matrix for latest 10 samples.
    \item Accept proposal $\boldsymbol{\beta}_k^{\gamma*}$ with probability $p=\min\left\{1, \varphi(\boldsymbol{\beta}_k^{\gamma*}\vert \cdots)/\varphi(\boldsymbol{\beta}_k^{\gamma^{(m-1)}}\vert \cdots)\right\}$, in which $\varphi(\boldsymbol{\beta}_k^{\gamma*}\vert \cdots)$ is the full conditional distribution for the corresponding basis coefficients given the data $\boldsymbol{Y}$ and the current values of all the remaining parameters:
    \begin{footnotesize}
    \begin{equation*}
        \varphi\left(\boldsymbol{\beta}_k^{\gamma^{(m-1)}}\big\vert \cdots\right)\propto \prod_{t=1950}^{2017}\prod_{j=1}^{\numStations}\varphi\left(Y_{tj}\big\vert R^{(m-1)}_t, \boldsymbol{W}^{(m-1)}_t,\delta^{(m-1)},\btheta^{(m-1)}_j,\tau^{2^{(m-1)}}\right)\times \prod_{i\in I_k} \phi\left(\frac{\beta_i^{\gamma^{(m-1)}}}{\sigma_{\beta}^{\gamma^{(m-1)}}}\right).
    \end{equation*}
    \end{footnotesize}
\end{enumerate}

\section{Empirical dependence strength}\label{sec:chi_emp}
In spatial extremes, the bivariate measure of upper tail dependence $\chi_{h}(u)$ is commonly used to describe the extremal dependence structure \citep{davison2012statistical, huser2019modeling}. Specifically, for two locations $\bs_i$ and $\bs_j$ with a distance $h>0$, the measure is defined by

\begin{equation*}
    \chi_{h}(u) = P\{F_j(X_j) > u \mid F_i(X_i) > u\} = \frac{P\{F_j(X_j) > u, F_i(X_i) > u\}}{P\{F_j(X_j) > u\}},
\end{equation*}
in which $X_i=X(\bs_i)$, $X_j=X(\bs_j)$ and $F_i$ and $F_j$ are their marginal distribution functions, respectively. Note that the statistical model in Section~\ref{sec:eva} assumes that  
$\{X(\bs):\bs\in \mathcal{D}\}$ has a stationary and isotropic dependence structure; see Section~\ref{sec:conc} for discussion on relaxing this assumption. 

\begin{figure}[!t]
\centering
\captionsetup[subfigure]{justification=centering}
    \begin{subfigure}[t]{.328\linewidth}
    \centering\includegraphics[height=1.06\linewidth]{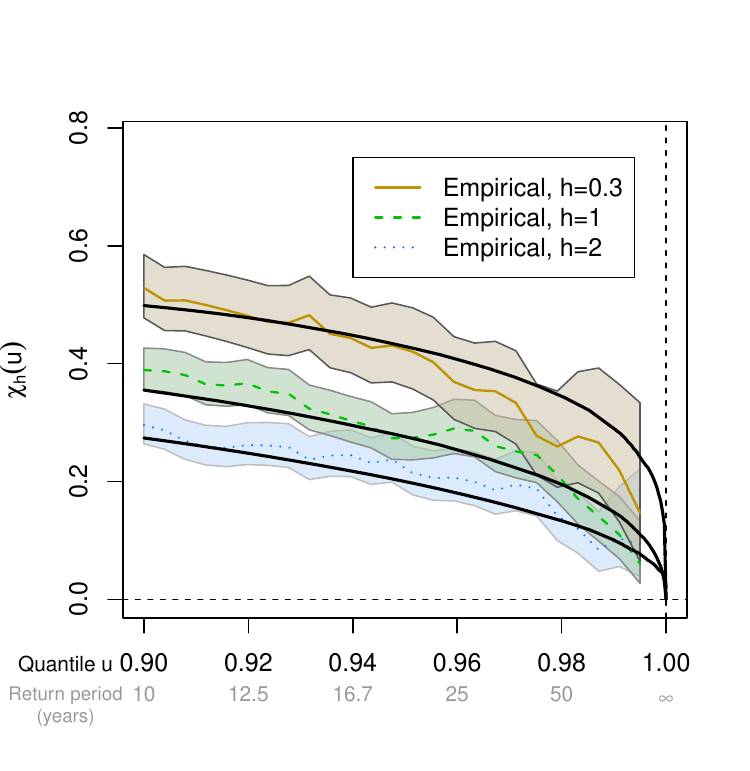}
    \caption{PNW region}
    \label{subfig:chi_emp}
    \end{subfigure}
    \begin{subfigure}[t]{.328\linewidth}
    \centering\includegraphics[height=1.06\linewidth]{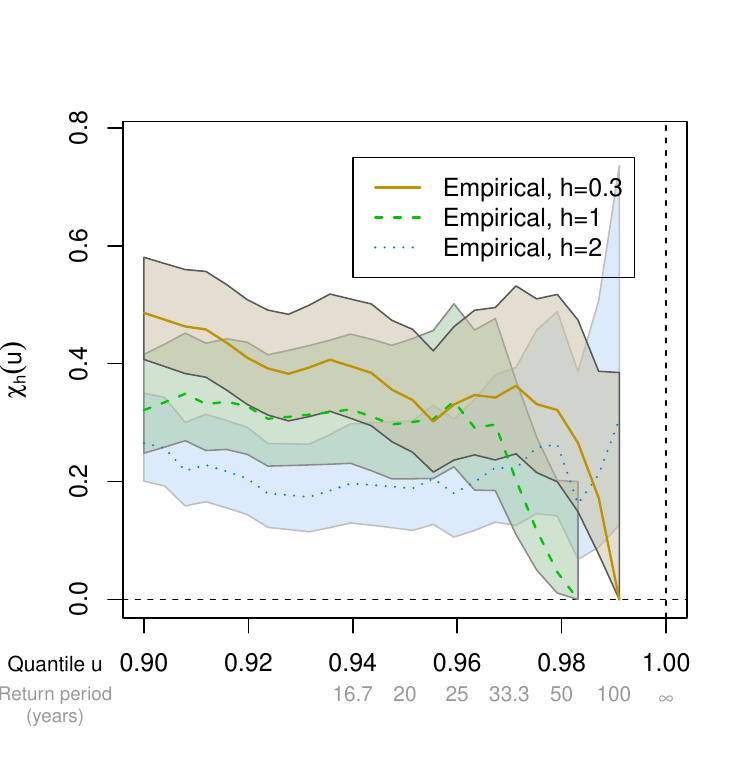}
    \caption{WWA region}
    \label{subfig:chi_emp_west}
    \end{subfigure}
    \caption{Empirical estimates of $\chi_h(u)$ for the 2021 TXx for inter-station distances of $h=0.3$ (approximately 30 km), $h=1$ (approximately 100 km) and $h=2$ (approximately 200 km) within the PNW region (left) and the WWA region (right); the 95\% confidence band is also shown for empirical estimates. The model-based estimates from the spatial analysis is only shown for the PNW region because the dependence parameter $\delta$ and the covariance parameters $\theta_W$ are imposed on the PNW region. 
    }
    \label{fig:chi_emp}
\end{figure}

To show how the measure $\chi_h(u)$ varies for the TXx over the PNW and WWA regions, Figure \ref{fig:chi_emp} shows the empirical and model-based (or parametric) estimates. Similarly to Figure 8 of \citet{zhang2021hierarchical}, we generated data-driven estimates of $\chi_h(u)$ empirically at distances of $h=0.3$ (approximately 30 km), $h=1$ (approximately 100 km) and $h=2$ (approximately 200 km) by collecting all pairs of GHCN stations whose locations are $h$ apart (within a small tolerance $\epsilon = 0.002$), and then computing the empirical estimate of $\chi_h(u)$ along with a 95\% nonparametric confidence envelope. This way of obtaining the nonparametric estimator of $\chi_{h}(u)$ resembles an empirical variogram estimator. For parametric estimates, we only overlay them on empirical estimates for the PNW region because our model is fit over this region. We take samples from the converged MCMC chain and use parameters from each iteration to generate the smooth $\chi_h(u)$ based on the statistical model described by Eq.~\eqref{eqn:noisyScalemix}. The black lines signify averages of these smooth $\chi_h(u)$ curves. The results in Figure \ref{fig:chi_emp} demonstrate that our copula model \eqref{eqn:noisyScalemix} captures the dependence structure of the transformed block maxima well in that the parametric estimates of $\chi_h(u)$ fall within the confidence bands of the empirical estimates.

On a side note, we notice that the empirical estimates of $\chi_h(u)$ within the WWA region have more uncertainty than those of the PNW region. This is mainly due to the less number of stations in a smaller region. However, $\chi_h(u)$ curve in WWA is on average lower than PNW at all three distances, indicating weaker dependence in the joint tail of the copula within WWA. This reveals a major limitation of our methodology --- one set of dependence parameter $\delta$ and covariance parameters $\theta_W$ governs the dependence structure of the entire domain, while the different subregions of a larger spatial domain may exhibit nonstationary dependence properties.

\end{document}